\documentclass[twocolumn]{aastex631}

\usepackage{amsmath}
\usepackage[T1]{fontenc}
\shorttitle{$\gamma$-rays from MADs in Radio Galaxies}
\shortauthors{Kuze et al.}
\graphicspath{{./}{figures/}}

\begin{document}

\title{High-energy Gamma-rays from Magnetically Arrested Disks in Nearby Radio Galaxies}

\correspondingauthor{Riku Kuze}
\email{r.kuze@astr.tohoku.ac.jp}

\author[0000-0002-5916-788X]{Riku Kuze}
\affiliation{Astronomical Institute, Graduate School of Science, Tohoku University, Sendai 980-8578, Japan}

\author[0000-0003-2579-7266]{Shigeo S. Kimura}
\affiliation{Frontier Research Institute for Interdisciplinary Sciences, Tohoku University, Sendai 980-8578, Japan}
\affiliation{Astronomical Institute, Graduate School of Science, Tohoku University, Sendai 980-8578, Japan}

\author[0000-0002-7114-6010]{Kenji Toma}
\affiliation{Frontier Research Institute for Interdisciplinary Sciences, Tohoku University, Sendai 980-8578, Japan}
\affiliation{Astronomical Institute, Graduate School of Science, Tohoku University, Sendai 980-8578, Japan}



\begin{abstract}
  The origins of the GeV gamma-rays from nearby radio galaxies are unknown. Hadronic emission from magnetically arrested disks (MADs) around central black holes (BHs) is proposed as a possible scenario. Particles are accelerated in the MAD by magnetic reconnection and stochastic turbulence acceleration. 
  We pick out the fifteen brightest radio galaxies in the GeV band from the Fermi 4LAC-DR2 catalog and apply the MAD model. We find that we can explain the data in the GeV bands by the MAD model if the accretion rate is lower than 0.1\% of the Eddington rate. For a higher accretion rate, GeV gamma-rays are absorbed by two-photon interaction due to copious low-energy photons. 
  If we assume another proposed prescription of the electron heating rate by magnetic reconnection, the MAD model fails to reproduce the GeV data for the majority of our sample. This indicates that the electron heating rate is crucial.
  We also apply the MAD model to Sgr A* and find that GeV gamma-rays observed at the Galactic center do not come from the MAD of Sgr A*. We estimate the cosmic ray intensity from Sgr A*, but it is too low to explain the high-energy cosmic ray intensity on Earth.
\end{abstract}



\section{Introduction} \label{sec:intro}

Radio-loud active galactic nuclei (AGN) have powerful relativistic jets that have a strong influence on star formation activities in host galaxies and thermodynamics of gases in galaxy clusters. These AGN also exhibit broadband non-thermal emission signatures from radio to GeV-TeV gamma-rays. However, the production mechanism and physical nature of the jets and the non-thermal emission are still unknown \citep[see e.g.,][for recent reviews]{Blandford2019,Hada2019}.

Blazars, a subclass of radio-loud AGN seen along the jet axis, provide a dominant contribution to the GeV gamma-ray sky \citep{Ackermann2015_DiffGammaray, Fermi4FGL2020}. Owing to the relativistic beaming effect, emission from the jets dominates over the other emission components. Their rapid variabilities also indicate that the gamma-ray emission site should be as compact as sub-pc scales \citep[e.g.,][]{Fermi_Mrk421_2011}. 

Radio galaxies, off-axis counterparts of blazars, are also detected in GeV-TeV gamma-rays \citep[e.g.,][]{InoueY2011,Stecker2019,MAGIC2020,HESS2020,deMenezes2020,Tomar2021}. The gamma-ray production sites for radio galaxies are controversial because relativistic beaming effects should be weaker in these objects. Leptonic compact jet models are actively discussed as a standard scenario \citep[e.g.,][]{Fermi_NGC1275_2009,MAGIC2020}, but at least in M87, this scenario failed to reproduce the magnetic field strength estimated by core-shift measurements in the radio bands \citep{Kino2015,Jiang2021}. If we assume the strong magnetic fields given by the radio observations, the resulting gamma-ray spectra are far below the observed flux \citep{EHT_MWL2021}. This motivates ones to investigate another scenarios, such as hadronic jets \citep{Reynoso2011,MAGIC2020}, large-scale jets \citep{Hardcastle2011}, hybrid jets \citep{Fraija2016}, and black-hole (BH) magnetospheres \citep{Hirotani2016,Kisaka2020}. However, all the scenarios have some difficulties or conflicts with other observations \citep[see][and references therein]{Kimura2020}.

\citet{Kimura2020} proposes hadronic emission in magnetically arrested disks (MADs; \citealt{Bisnovatyi-Kogan1974,Narayan2003}) as an alternative scenario. Owing to their strong magnetic fields, MADs can launch powerful relativistic jets via Blandford-Znajek mechanism \citep{Tchekhovskoy2011,Mckinney2012,EHT2019v}. Thus, the presence of jets implies the existence of strong magnetic fields in the vicinity of the BH, which suggests that reconnection-driven particle acceleration \citep{Hoshino2012, Guo2020} taking place in the MAD is important. The accelerated protons emit GeV gamma-rays via the synchrotron process. This model can reproduce the GeV-TeV gamma-ray data from M87 and NGC 315, but the majority of the GeV-detected radio galaxies are unexplored yet.

The existence of non-thermal particles in accretion flows is supported in terms of both theories and observations. Recent general relativistic magnetohydrodynamic (GRMHD) simulations revealed that MADs can induce magnetic reconnection in highly magnetized plasmas with the magnetization parameter of $\sigma=B^2/(4\pi m_p n_p c^2)\gtrsim1$ \citep{Ball2018,Ripperda2020,Ripperda2022}. These reconnection events very efficiently produce non-thermal particles, according to particle-in-cell simulations \citep{Zenitani2001,Guo2016,Zhang2021}. Also, accretion flows are turbulent, under which stochastic acceleration process may produce non-thermal particles efficiently \citep{Kimura2016,Comisso2018,Zhdankin2018,KimuraTomida2019}.

The multi-wavelength and multi-messenger observations also provide hints of non-thermal signatures in accretion flows. \citet{IceCube2020ps_search} reported a $\sim3\sigma$ high-energy neutrino signal from  NGC 1068, a nearby X-ray bright Seyfert galaxy. This motivates ones to consider non-thermal hadronic emissions in accretion flows \citep{InoueY2019,Murase2020,Gutierrez2021,Kheirandish2021,KimuraMurase2021}. GeV gamma-ray detections are also reported from radio-quiet AGN \citep{wojaczynski2015,Fermi4FGL2020}, indicating non-thermal activity in accretion flows. The flaring activities of Sgr A* in infrared and X-rays are also considered as the non-thermal phenomena triggered by magnetic reconnection \citep{Dexter2020,Porth2021,Gravity2021}. 

In this paper, we investigate the characteristics of radio galaxies that can be explained by the MAD model by applying the model to fifteen GeV-loud radio galaxies.
We also apply our MAD model to Sgr A* to see whether gamma-rays from the Galactic center can originate from the accretion flow. This paper is organized as follows. In Section \ref{sec:model}, we describe the MAD model constructed by \cite{Kimura2020}. In Section \ref{sec:results}, we classify the radio galaxies by comparing the calculated photon spectra to the gamma-ray data and discuss the characteristics of radio galaxies. We also examine another prescription of the electron heating rate. In Section \ref{sec:SgrA}, we also apply the MAD model to Sgr A* and discuss Sgr A* as a cosmic ray (CR) source. In Section \ref{sec:summary}, we present our conclusions.

\section{MAD model} \label{sec:model}

We calculate the photon spectra with the MAD model constructed by \cite{Kimura2020}.
In this model, particles are accelerated by the magnetic reconnection at the edge of the accretion disk \citep{Ball2018,Ripperda2020} and the turbulence in the accretion disk \citep{Yuan2003,Kimura2016,KimuraTomida2019}. 
We consider that plasma is accreted onto a supermassive BH of mass $M$.
The mass accretion rate, $\dot{M}$, and the size of emission region, $R$, are normalized by the Eddington rate and by the gravitational radius, respectively, i.e., $\dot{M}c^2 = \dot{m}L_{\rm{Edd}}$ and $R=\mathcal{R}R_G = \mathcal{R}GM/c^2$. We use the notation of $Q_{\rm x}={Q}/{10^{X}}$ in cgs units, except for the BH mass, $M$ ($M_9 = M/[10^9 M_\odot]$).
This model considers the emission by thermal electrons, non-thermal electrons, non-thermal protons, and secondary electron-positron pairs produced by the Bethe-Heitler process ($p+\gamma \rightarrow p + e^+ + e^-$) and the two-photon interaction ($\gamma + \gamma \rightarrow e^+ + e^-$).

We determine the electron temperature by balancing the electron heating rate with the cooling rate (see Appendix \ref{temax}). 
The electron heating mechanism in the MAD has not been established yet \citep{Rowan2017, Kawazura2019}. 
We consider that magnetic reconnection is the dominant electron heating mechanism and use the formalism of \cite{Hoshino2018} as a fiducial prescription. Then, the electron heating rate is given by
\begin{equation}
 {Q_e} = \left( \frac{m_eT_e}{m_pT_p}\right)^{1/4} Q_p,
\label{eqthmlele_1}
\end{equation}
where $m_e$ and $m_p$ are the mass of an electron and a proton, respectively,  $Q_p = \epsilon_{\rm NT} \epsilon_{\rm dis} \dot{M}c^2$ is the proton heating rate, $\epsilon_{\rm NT}$ is the fraction of the non-thermal particle energy production rate to the dissipation rate, and $\epsilon_{\rm{dis}}$ is the fraction of the dissipation rate to the accretion rate.
The thermal energy of electrons is lost by radiation cooling or advection to the BH. 
For $\dot{m}$ higher than $\dot{m}_{\rm crit}$ given in Appendix \ref{temax}, the radiation cooling balances the heating rate because the cooling rate is efficient owing to the high density and strong magnetic field.
For $\dot{m}$ lower than $\dot{m}_{\rm{crit}}$, the thermal electrons fall to the central BH before they cool. In this case, the electron temperature is estimated as $T_e/T_p \simeq Q_e/Q_p$ (see Appendix \ref{temax}) \citep[see also][]{KimuraKashiyamaHotokezaka2021}. 

As the escape process, we consider the infall to the BH and the diffusion. The infall timescale is $t_{\rm{fall}} \approx R/V_R$, where $V_R = \alpha V_K /2$ is the radial velocity, $\alpha$ is the viscous parameter \citep{Shakura1973}, and $V_K = \sqrt{GM/R}$ is the Kepler velocity. The diffusion timescale is $t_{\rm{diff}} \approx R^2/D_R$, where $D_R \approx \eta r_{i,L}c/3$ is the diffusion coefficient, $r_{i,L} = {E_i}/{(eB)}$ is the Larmor radius, $\eta r_{i,L}$ is the effective mean free path, $\eta$ is the numerical factor, $B = \sqrt{8\pi \rho C_s^2/\beta}$ is the magnetic field, $\rho = \dot{M}/(4\pi R H V_R)$ is the mass density, $H \sim (C_s/V_K)R$ is the scale height of an accretion disk, $C_s \approx V_K/2$ is the sound speed, and $\beta$ is the plasma beta.

We phenomenologically estimate the acceleration timescale as 
\begin{equation}
t_{\rm{acc}} \approx \eta \frac{r_{i,L}}{c} \left( \frac{c}{V_A} \right)^2,     
\label{eq_acc}
\end{equation}
where $V_A = {B}/{\sqrt{4\pi \rho}}$ is the Alfv\'en velocity. 
We consider only the synchrotron cooling as the cooling process of primary electrons and secondary electron-positron pairs because the other processes are negligible. We consider the proton synchrotron, $pp$ collision ($p+p \rightarrow p+p+\pi$), photomeson production ($p+\gamma \rightarrow p+\pi$), and the Bethe-Heitler process as the protons cooling processes. 
In the range of our investigation, $pp$ collision and photomeson production are inefficient because of the low number density of thermal protons and high threshold energy for photomeson production than the Bethe-Heitler process. 

High-energy protons and photons interact with the low-energy photons. In \citet{Kimura2020}, since they consider the low $\dot{m}$ radio galaxies, the photons by the thermal electrons are dominant as the target photons for the Bethe-Heitler process and the two-photon interaction. For a high $\dot{m}$, the number density of photons produced by non-thermal particles is comparable to or higher than that produced by thermal electrons, and thus, we take into account all the photons inside the MAD as the target photons for the two-photon interaction and the Bethe-Heitler process. We iteratively calculate the photon and the electron-positron pairs spectra until converged.

\section{Results for Radio Galaxies} \label{sec:results}
\subsection{Properties of the MAD Model} \label{subsec:modelproperty}
\begin{figure}[tb]
 \includegraphics[width = 0.5\textwidth]{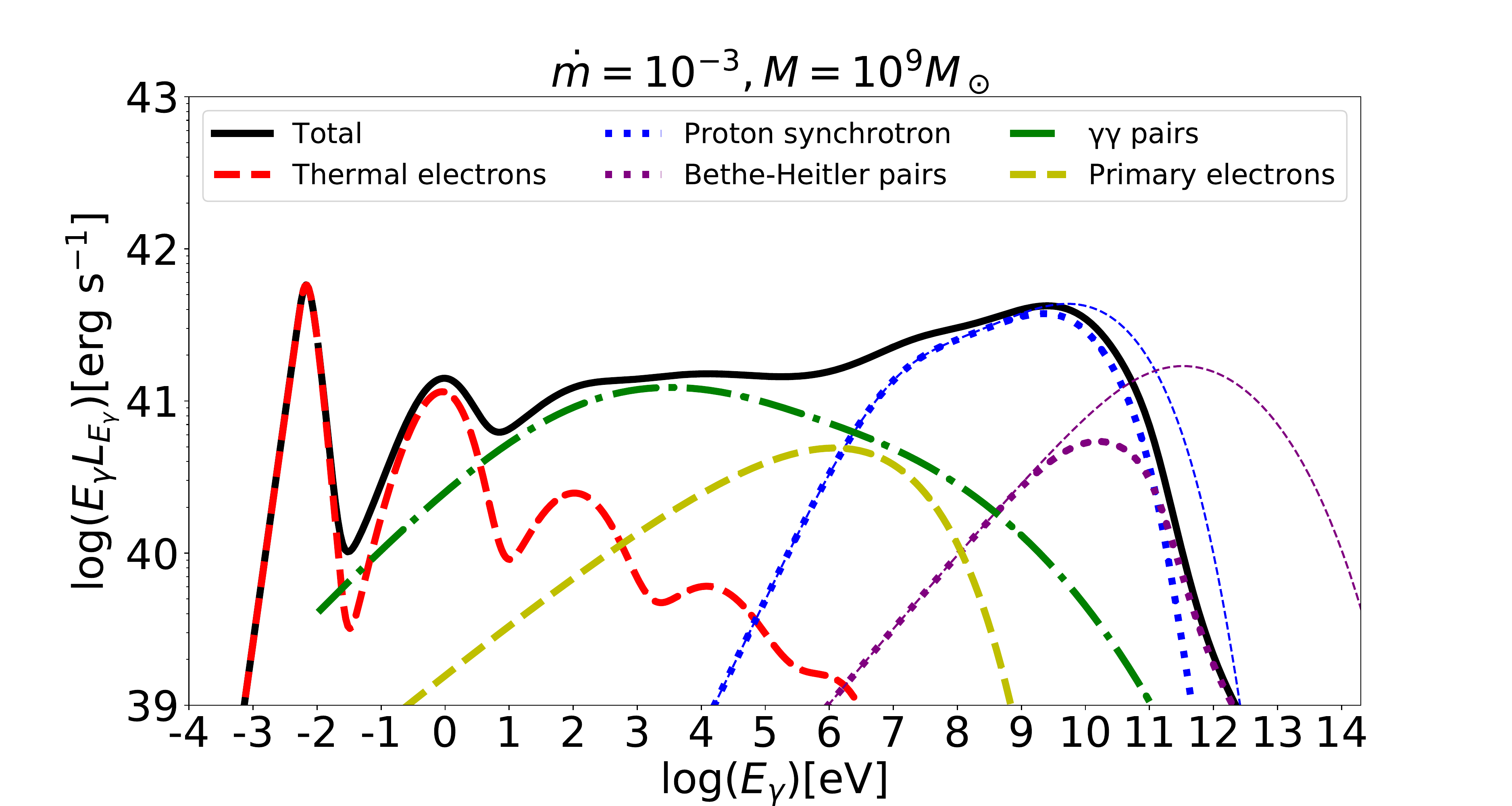}   
 \caption{The typical broadband photon spectrum by the MAD model for $\dot{m} = 10^{-3} \ {\rm{and}}\ M = 10^9 M_\odot$. The thick and thin lines are the photon spectra after and before internal attenuation by the two-photon interaction, respectively. The black-solid, red-dashed, green-dot-dashed, yellow-dashed, blue-dotted, and purple-dotted lines are the total luminosity by the MAD model, thermal electrons synchrotron and Comptonization, synchrotron by the secondary electron-positron pairs by the two-photon interaction, primary electrons synchrotron, primary protons synchrotron, and synchrotron by the secondary electron-positron pairs by the Bethe-Heitler process, respectively. }
 \label{fig_typ}
\end{figure}
\begin{table}[tbp]
  \caption{Our fiducial parameter set, which is the same as that used in \citealp{Kimura2020}.}
  \begin{tabular}{ccccccc} \hline
    $\mathcal{R}$ & $\alpha$ & $\beta$ & $\epsilon_{\rm{NT}}$ & $\epsilon_{\rm{dis}}$ & $\eta$ & $s_{\rm{inj}}$\\ \hline
    10 & 0.3 & 0.1 & $0.33$ & 0.15 & 5 &1.3 \\ \hline
   \end{tabular}
  \label{tb_para}
\end{table}

We show typical photon spectrum by the MAD model  for $M = 10^9 M_\odot$ and $\dot{m}=10^{-3}$ in Figure \ref{fig_typ}. The other parameter values are the same as \cite{Kimura2020} (see Table \ref{tb_para}), by which the spectra for M87 and NGC 315 are explained. 
The thick and thin lines are the photon spectra after and before internal attenuation by the two-photon interaction, respectively. The non-thermal protons emit GeV gamma-rays (blue-dotted line), the secondary electron-positron pairs by the Bethe-Heitler process emit TeV gamma-rays (purple-dotted line), and the primary electrons and the secondary electron-positron pairs by the two-photon interaction emit X-rays  (yellow-dashed and green-dot-dashed lines, respectively) as seen in Figure \ref{fig_typ}. 

Both of the most efficient energy loss timescales of non-thermal protons at the highest energy range, $t_{\rm syn}$ and $t_{\rm diff}$, have the same dependence $\propto E_p^{-1}$, so that either of the energy loss processes dominates over the other in the entire proton energy range. The equality $t_{\rm syn} = t_{\rm diff}$ gives the critical mass,
\begin{align}
 M_{\rm{crit}} &= \frac{\sigma_T c^4 m_p^5 }{128 \pi G e^2 m_e^4} \dot{m}^{-3} \mathcal{R}^{7/2} \alpha^{3} \beta^{3} \eta^2 \nonumber \\
 &\approx 8.3 \times 10^{5} \dot{m}_{-3}^{-3} \mathcal{R}_1^{7/2} \alpha_{-0.5}^{3} \beta_{-1}^{3} \eta_{0.5}^{2} \ M_\odot .\label{eq_Mcrit}
\end{align} 
The synchrotron cooling is dominant if $M > M_{\rm{crit}}$ for a given $\dot{m}$, while the diffusion loss is dominant if $M < M_{\rm{crit}}$. 
For a fixed value of $M$, the magnetic field is stronger for higher $\dot{m}$, which leads to higher synchrotron power. The diffusive escape timescale is longer for higher $\dot{m}$ due to a smaller Larmor radius.
For the example shown in Figure \ref{fig_typ}, the synchrotron cooling is dominant, i.e., $M>M_{\rm crit}$. We find that it is also the case for vast majority of radio galaxies in our sample (see Figures \ref{fig_ma} and \ref{fig_ma_chael}).

The analytical estimates of the peak energy and luminosity of the proton synchrotron spectrum in the synchrotron cooling case are given as follows.
Because of the hard spectral index of protons, the proton synchrotron spectrum has a peak at the synchrotron frequency for $E_p = E_{p,{\rm{cut}}}$. Balancing the synchrotron cooling and acceleration timescales, we obtain $E_{p,{\rm{cut}}}$ as
\begin{align}
    E_{p,{\rm{cut}}} &= \sqrt{\frac{6 \pi e}{\eta \sigma_T B}} c V_A \frac{m_p^2}{m_e} \nonumber \\
    &\approx  5.7\times10^9 \dot{m}^{1/4}_{-3} M^{1/4}_{9} \mathcal{R}^{1/8}_{1} \alpha^{1/4}_{-0.5} \beta^{-1/4}_{-1} \eta^{-1/2}_{0.5} \ {\rm{GeV}}.
\end{align}
We obtain the peak frequency of the synchrotron spectrum by the non-thermal protons as
\begin{align}
    E_{\gamma,p,{\rm{peak}}} &= \frac{3e^2 h}{\sigma_T m_p c \eta}\left( \frac{V_A}{c}\right)^2 \left( \frac{m_p}{m_e}\right)^2 \nonumber \\
    &\approx 46 \mathcal{R}_1^{-1} \beta_{-1}^{-1} \eta_{0.5}^{-1} \ {\rm{GeV}}.
\label{eqpeak1}
\end{align}
Since the synchrotron cooling is the dominant energy loss timescale, we can approximate that all the energies used for non-thermal proton acceleration are converted to the synchrotron photon energy. Then, the photon luminosity for the proton synchrotron process is estimated to be
\begin{equation}
    L_{\gamma,{\rm{psyn}}} \approx 4.0 \times 10^{42} \dot{m}_{-3} M_9 \epsilon_{\rm{NT}_{-0.5}} \epsilon_{\rm{dis}_{-1}} \ {\rm{erg \ s^{-1}}}.    
    \label{eqlumpeak}
\end{equation}
We should note that this estimate provides the integrated photon luminosity. The differential photon luminosity given in Figure \ref{fig_typ} is lower than $L_{\gamma,\rm psyn}$ because of the bolometric correction.

\subsection{Application to the various Radio Galaxies} \label{sec:classification}
We search for bright radio galaxies in the GeV gamma-ray band from the Fermi 4LAC-DR2 catalog \citep{Ajello2020}. We pick up the fifteen brightest objects after excluding Fornax A, M87, and NGC 315. We exclude the Fornax A because the emission region of gamma-rays is extended and the contribution of the core is lower than 18\% \citep{Ackermann2016}. We also omit M87 and NGC 315 since these objects are already explained in \cite{Kimura2020}. For Cen A, we use the gamma-ray data of the core while the extended component is also observed. In particular, 
the HESS data ($E_\gamma >300$ GeV) should be from the extended component \citep{HESS2020}. Then the 100 GeV data should be the sum of the jet component and disk component.
The theoretical model for the HESS data predicts that the extended jet contributes to the 100 GeV data very marginally \citep{HESS2020}. The MAD contribution to the 100 GeV data is uncertain. Thus,
we use the 2-20 GeV data for the fitting procedure and restrict the MAD model not to exceed the data above 100 GeV.

We compare the spectra obtained by the MAD model to the observed ones. To evaluate the goodness of fit, we use $\chi^2$ method. $\chi^2$ is the quantity written as
\begin{equation}
 \chi^2 = \sum_i \left( \frac{{F_{{\rm{data}},i}} - { F_{{\rm{model}},i}}}{\sigma_i} \right)^2,
 \label{chi2}
\end{equation}
where $i$ represents the observational data points, $F_{{\rm data},i}$ is the gamma-ray flux data, $F_{{\rm model},i}$ is the calculated gamma-ray flux, and $\sigma_i$ is the observational error. In this calculation, we use only the gamma-ray data and change only $\dot{m}$ in the parameters. We consider that the GeV data are explained by the MAD model if $Q\geq0.01$, where $Q$ is the probability that $\chi^2$ exceeds the obtained value by Equation (\ref{chi2}). 
We consider that the emission from the jet predominantly contributes to the lower-energy data. The photon flux from radio galaxies shows some variability in all the energy bands, and we regard them as the jet contribution. Thus,  the contribution by the MAD model should be below the lowest data points in radio to X-ray bands.

We classify the results into three; Excellent, Good, and Bad. We classify objects into Excellent if we can explain the gamma-ray data with the parameters in Table \ref{tb_para} and the cataloged value of $M$. 
We show the values of $M$, distance from Earth, $\dot{m}$, $\chi^2/\nu$ , and $Q$ for the Excellent objects in Appendix \ref{ap_hoshino}, where $\nu = N-m$ is the degree of freedom, $N$ is the number of the data, and $m$ is the number of the changing parameters. Since we only change $\dot{m}$, we set $m=1$. We also show the photon spectra of these objects in Appendix \ref{ap_hoshino}. We find that the accretion rates of all the Excellent objects are less than $10^{-3}$. 

For some objects, it is hard to explain the gamma-ray data with the parameters in Table \ref{tb_para} and the cataloged value of $M$.
This is because the GeV gamma-rays have the cut-off by the two-photon interaction. In order to achieve the high GeV gamma-ray flux, 
we may use higher values of $\mathcal{R}$ and $M$. The uncertainty of $M$ is about a factor of 3 \citep[e.g.,][]{Kormendy2013}. We classify objects into Good if we can explain the GeV data with $\mathcal{R}=30$ and $M$ three times higher than the cataloged value.
We only change $\dot{m}$ during the fitting procedure. 
Thus, we calculate $Q$ with $m = 1$. We show the resulting quantities and the photon spectra for the Good objects in Appendix \ref{ap_hoshino}. The accretion rates of the Good objects are around $10^{-3}$. 
Owing to the larger emission region, absorption by the two-photon interaction is suppressed, which enables the MAD model to explain GeV data for $\dot{m}\gtrsim10^{-3}$.

The other objects are classified as Bad. We show the photon spectra for the Bad objects and the quantities for these spectra in Appendix \ref{ap_hoshino}. 
There are two types of Bad objects. 
One type has a cut-off due to the two-photon interaction below the GeV energy, which leads to a mismatch in the multi-GeV data. 
The other type has luminous synchrotron emission by the secondary electron-positron pairs by the two-photon interaction, which overshoots the X-ray data.

To see the features of the radio galaxies, we plot $M$ and $\dot{m}$ for the objects of the three classes in Figure \ref{fig_ma},
where values of $M$ and $\dot{m}$ for individual objects are tabulated in tables in Appendix \ref{ap_hoshino}.
As can be seen, we can explain the gamma-ray data by the MAD model if $\dot{m}$ is lower than $10^{-3}$. The number density of low-energy photons is higher for a higher $\dot{m}$, and then, the two-photon interaction is more efficient. Thus, the photon spectra by the MAD model have the cut-off below the GeV range, and we cannot explain the gamma-ray data for a higher $\dot{m}$. For the jet model, GeV gamma-ray absorption is inefficient owing to the large emission region, and thus, we consider that the GeV gamma-rays come from the jet for high $\dot{m}$ radio galaxies.

\begin{figure}[tb]
 \includegraphics[width = 0.5\textwidth]{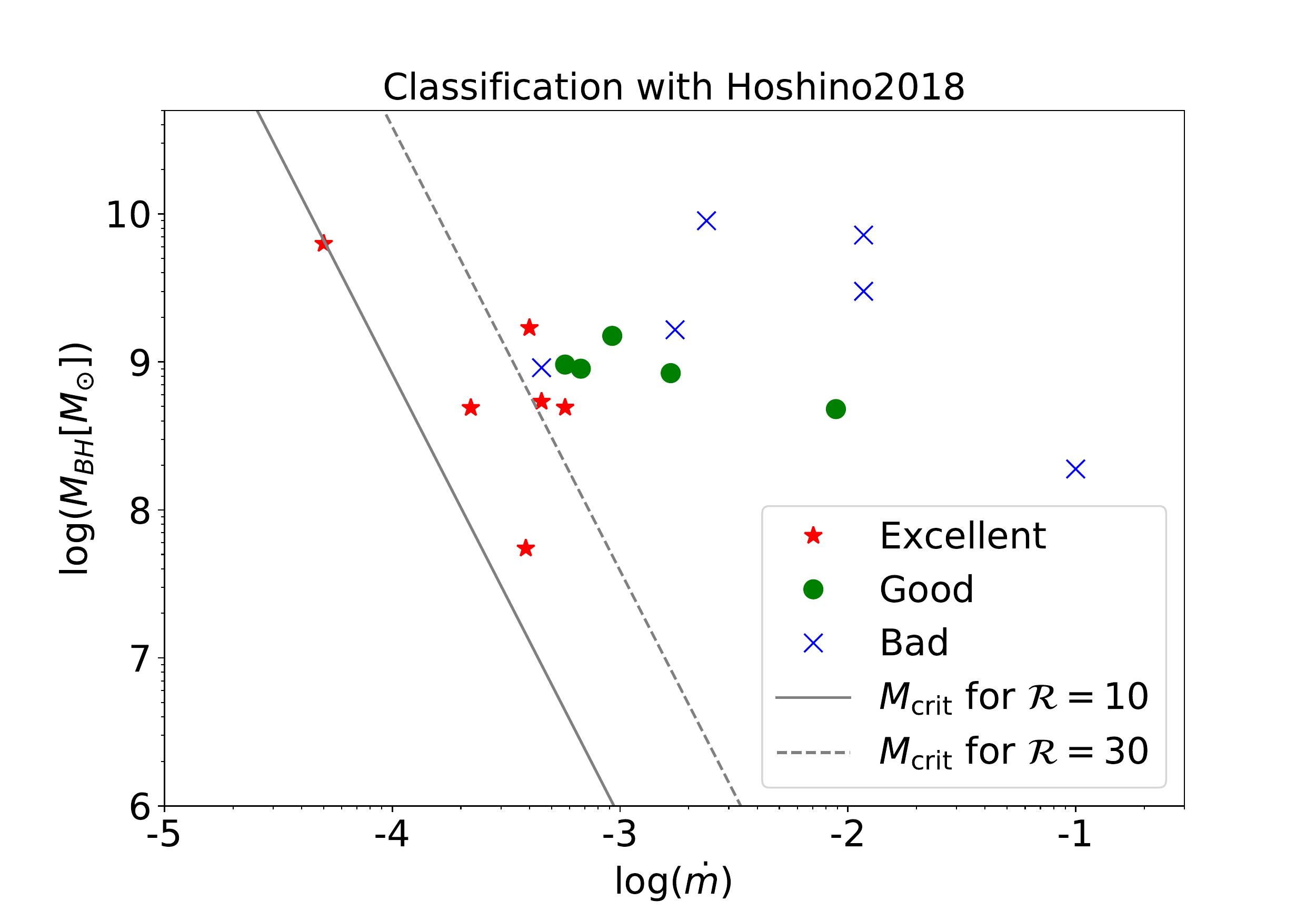}   
 \caption{Scatter plot of our sample in $M-\dot{m}$ plane. 
The red-star, green-circle, and blue-cross indicate the Excellent, Good, and Bad objects, respectively.
The solid and dashed lines indicate $M_{\rm crit}$ given by Equation (\ref{eq_Mcrit}) for $\mathcal{R} = 10$ and $\mathcal{R} = 30$, respectively.
 }
 \label{fig_ma}
\end{figure}

\subsection{Another Formalism of the Electron Heating Rate}
\begin{figure}[tb]
 \includegraphics[width = 0.5\textwidth]{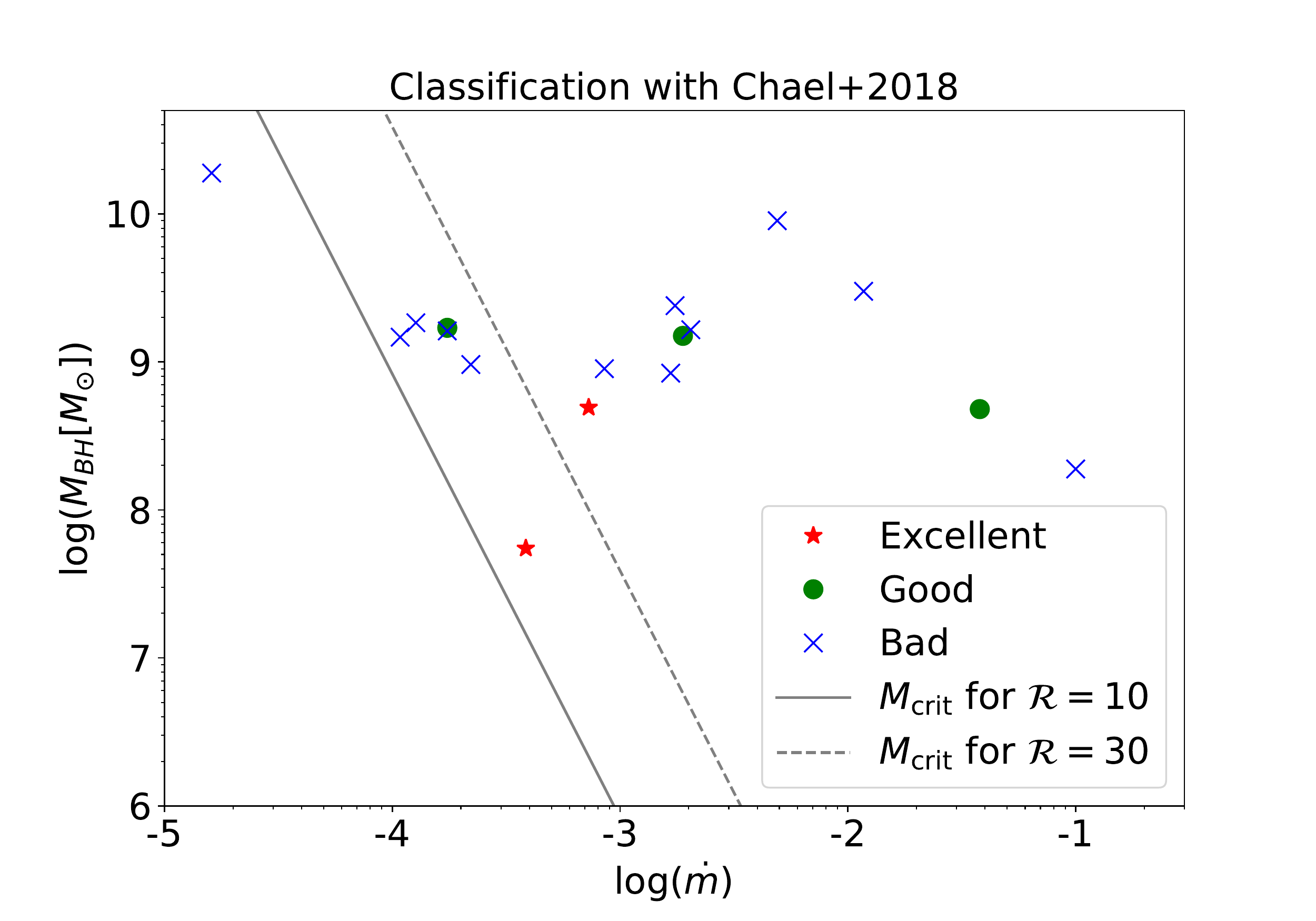}   
 \caption{Same as Figure \ref{fig_ma}, but with the electron heating rate given by \cite{Chael2018}.}
 \label{fig_ma_chael}
\end{figure}

The electron heating rate by magnetic reconnection has not been established yet. We also examine another prescription of the electron heating rate given by \cite{Chael2018},
\begin{equation}
    \frac{Q_e}{Q_p} = \frac{1}{2} \exp \left[ \frac{-(1-\beta/\beta_{\rm{max}})}{0.8 + \sigma^{1/2}} \right],
    \label{eqchael1}
\end{equation}
where $\beta_{\rm{max}} = 1/(4 \sigma)$. We show the photon spectra and the resulting quantities of the objects in Appendix \ref{chael}. We calculate the photon spectra for all the objects with this electron heating rate and classify them 
as we have done in Section \ref{sec:classification} by changing $\dot{m}$ with the same parameter set.
The classification results are shown in Figure \ref{fig_ma_chael}, where we see that all the classes (Excellent, Good, Bad) equally scatter in the $M$-$\dot{m}$ plane.
We find that $Q_e/Q_p\sim0.3$ if we use Equation (\ref{eqchael1}) with the parameters in Table \ref{tb_para}. On the other hand, $Q_e/Q_p\sim0.07$ by Equation (\ref{eqthmlele_1}).
The value of $Q_e/Q_p$ corresponds to the luminosity of the electrons, and thus, the luminosities in radio and X-ray bands are high if we use Equation (\ref{eqchael1}). This causes the model flux to overshoot the radio and X-ray data if we adjust $\dot{m}$ so that the resulting gamma-ray spectra match the GeV data. 
Equation (\ref{eqchael1}) leads to $0.2 < Q_e/Q_p < 0.4$ for $ 5 \leq r \leq 30$ and $0.01 \leq \beta \leq 1$. Thus, we cannot reconcile the results in Section \ref{sec:classification} even with a different parameter set. This indicates that the electron heating rate is crucial to explain the gamma-ray data by the MAD model.

\section{Sgr A*} \label{sec:SgrA}
\begin{figure}[tb]
 \includegraphics[width = 0.5\textwidth]{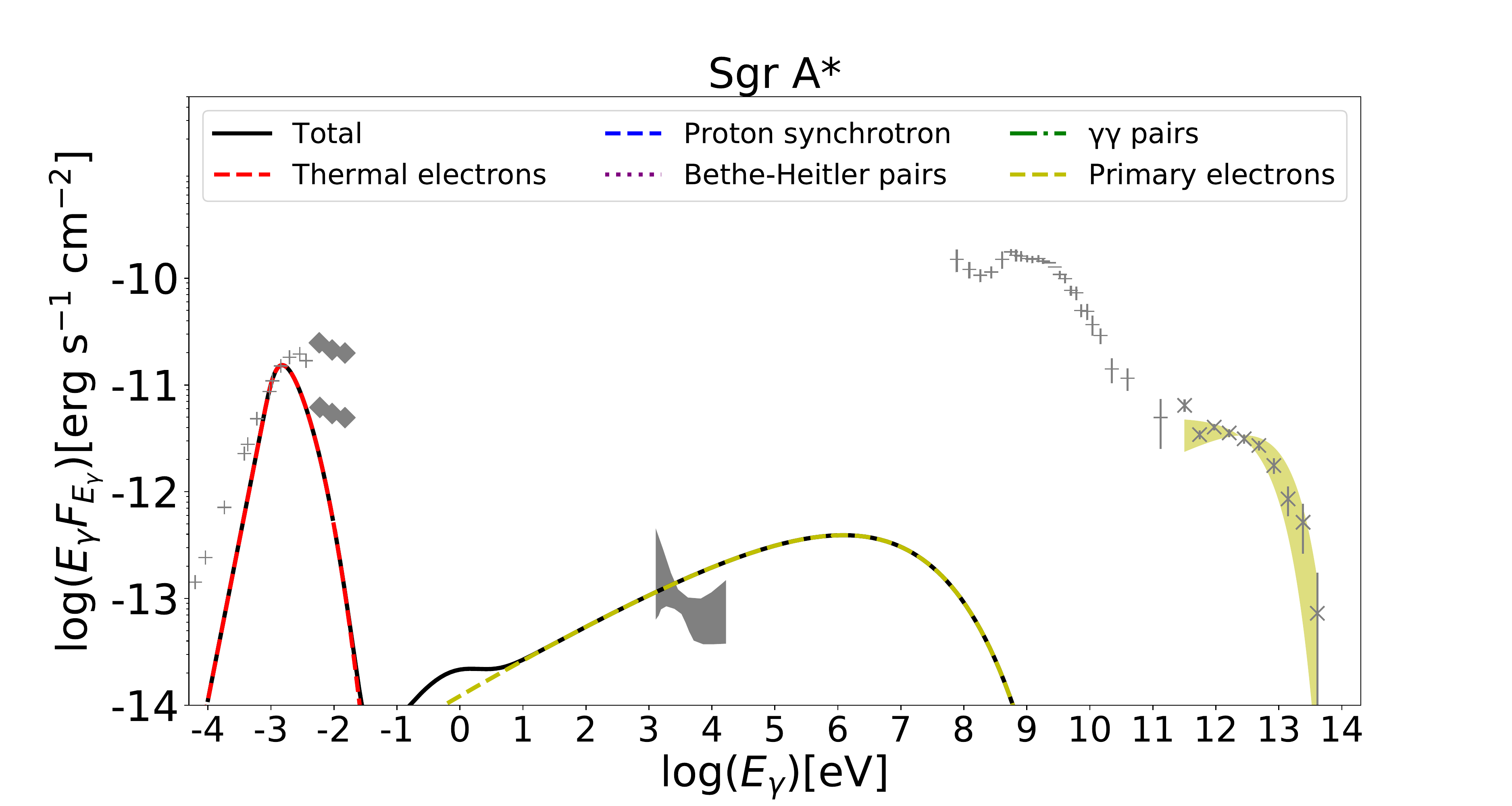}   
 \caption{ Photon spectrum fit to the Sgr A* data. The line types are the same as Figure \ref{fig_typ}. Data points are taken from \cite{Gravity2020} and \cite{Ahnen2017}. }
 \label{fig_sgr}
\end{figure}

Observations in the radio and X-ray bands imply that Sgr A* at the Galactic center has a hot accretion flow \citep{NarayanYiMahadevan1995, Manmoto1997, Yuan2003, Gravity2021}. Sgr A* is thought to have a MAD because the wind accretion by Wolf-Rayet stars can provide sufficiently large-scale magnetic flux \citep{Ressler2020}. A MAD is also expected to be formed in a low $\dot{m}$ system \citep{Kimura2021}, and Sgr A* is known to be a very low accretor. According to the observations by Event Horizon Telescope Collaboration, the time variability suggests a weakly magnetized accretion disk, but the other constraints favor a MAD \citep{EHT2022SgrA_I, EHT2022SgrA_V}. Here, we apply the MAD model to Sgr A*. We show the parameters in Table \ref{tb_sgr} and the photon spectrum in Figure \ref{fig_sgr}. For Sgr A*, $\epsilon_{\rm{NT}}$ needs to be much lower than that for the other radio galaxies to match the radio and X-ray data. We also find that $\dot{m}$ is too low to explain the GeV-TeV gamma-ray data.
For a lower $\dot{m}$, the diffusion timescale is much shorter than the synchrotron cooling timescale. Consequently, the radiative efficiency of the non-thermal protons is low. 
We cannot explain the GeV-TeV data even with $\epsilon_{\rm{NT}} = 0.5$ if we adjust $\dot{m}$ to reproduce radio data and ignore the X-ray data. 
As long as we use the same value of $\epsilon_{\rm NT}$ for electrons and protons, it is difficult to reproduce the GeV-TeV data and low-energy (radio to X-ray) data simultaneously.
The angular resolution of the GeV-TeV gamma-ray observation is about 0.1 degrees \footnote{\url{https://fermi.gsfc.nasa.gov/science/instruments/table1-1.html}}. 
This corresponds to 200 pc for the length scale at the Galactic center,
within which many other GeV-TeV gamma-ray source candidates exist.
We consider that other accretion models cannot explain GeV-TeV data because the $\epsilon_{\rm{NT}} = 0.5$ of the MAD model is close to the theoretical upper limit, and thus, we conclude that the sources of GeV-TeV gamma-rays are other objects in the Galactic center region.
\begin{deluxetable}{ccccccc}
\tablenum{2}
\tablecaption{BH mass, distance, accretion rate, $\mathcal{R}$, and $\epsilon_{\rm{NT}}$ for Sgr A*. \label{tb_sgr}}
\tablewidth{10pt}
\tablehead{
\colhead{Mass $[M_\odot]$}  & \multicolumn{0}{c}{Distance [kpc]} & \colhead{$\dot{m}$} & \colhead{$\mathcal{R}$} & \colhead{$\epsilon_{\rm{NT}}$}
}
\startdata
   $4.3 \times 10^{6}$ & 8.2 & $6 \times 10^{-7}$ & 10 & 0.007\\ 
\enddata
\tablecomments{The references for BH masses, distances are \cite{Gillessen2017}.}
\end{deluxetable}

Recent experiments show the distribution of CRs is anisotropic, and Galactic CRs of higher energies ($E_p \gtrsim 300 \ {\rm{TeV}} $) come from the direction of the Galactic center \citep{IceTop2013,Amenomori2017}. We investigate the CR intensity produced at Sgr A*. The luminosity of the CRs injected from the accretion disk of Sgr A* is approximated as $L_p$. This is because the diffusion timescale is much shorter than the synchrotron cooling timescale for Sgr A*. In \cite{KimuraMurase2018}, the CR luminosity escape from the galaxy is given by $L_{\rm{esc}}(E_p) = E_p U_{E_p}c {M_{\rm{gas}}}/{X_{\rm{esc}}}$, where $U_{E_p}$ is the differential energy density of the CRs, $M_{\rm{gas}} \sim 10^{10} M_{\odot}$ is the mass of the gas inside the Galaxy, $X_{\rm{esc}} \cong 8.7 r_1^{-1/3} \ \rm{g \ cm^{-2}}$ is the grammage, and $r_1 = (E_p/e)/(10\ \rm{GV})$. Assuming the steady state, we estimate the $E_p U_{E_p}$ from the balance of the CRs injected to and the escape from the interstellar medium. The cut-off energy for Sgr A* is $E_{p,{\rm{cut}}} \approx 2.5 \times 10^7 \ \rm{GeV}$.
For $E_p = E_{p,{\rm{cut}}}$, we estimate the CR intensity as
\begin{align}
    E_p^2\frac{dN}{dE_p dt dS d\Omega} &= \frac{c}{4 \pi} E_p U_{E_p} \\ \nonumber
    &\approx 2.1 \times 10^{-8} \ {\rm{GeV \ s^{-1} \ cm^{-2} \ sr^{-1}}}.
    \label{eqCR4}
\end{align}
The CR intensity obtained by the CR experiments is $1.2 \times 10^{-5}\ {\rm{GeV \ s^{-1} \ cm^{-2} \ sr^{-1}}} \ {\rm{at}} \ E_p = 2.8\times 10^7 \ {\rm{GeV}}$ \citep{Amenomori2008}. Thus, the contribution by Sgr A* is too low with the current activity.
 
Sgr A* is expected to be more active hundreds of years ago \citep{Koyama1996, Murakami2000} and may produce a larger amount of CRs that can explain TeV gamma-ray from the Galactic center region \citep{FujitaKimuraMurase2015,HESS2016}.
The activity of Sgr A* around 10 Myr ago may create the Fermi and eROSITA bubbles (see \citealt{Su2010} for the Fermi observation and \citealt{Predehl2020} for the eROSITA bubbles; see \citealp[e.g.,][]{Mou2014,Sarkar2017,Yang2022} for theoretical models).
If this activity also produces CRs efficiently, Sgr A* can explain the CR intensity of the present-day around the Knee observed on Earth \citep{FujitaMuraseKimura2017}.
If the past activities are in the MAD state, CRs can be accelerated to higher energies with an enhanced production rate. This may account for the light-mass galactic CRs reported in \cite{buitink2016}

\section{Conclusion} \label{sec:summary}
We statistically investigate the features of radio galaxies explained by the MAD model constructed by \cite{Kimura2020}.
We apply this model to the fifteen brightest GeV-loud radio galaxies picked out from the Fermi 4LAC-DR2 catalog. We classify these objects into three; Excellent, Good, and Bad, by comparing the spectra by the MAD model to the gamma-ray data. We find that we can explain the gamma-ray data by the MAD model if the accretion rate is lower than 0.1\% of the Eddington rate, while it is challenging to reproduce gamma-ray data for high $\dot{m}$ objects (see Figure \ref{fig_ma}).
For $\dot{m}\gtrsim10^{-3}$, the number density of the low-energy photons is so high that GeV gamma-rays cannot escape from the system due to efficient two-photon interactions. In this case, we consider that the GeV gamma-rays come from the jet rather than the disk because GeV gamma-ray absorption by the two-photon interaction is inefficient owing to the large emission region for the jet model. 

For the Bad objects, we cannot reproduce the GeV gamma-rays by the MAD model, but their accretion disks could be in the MAD states. For $\dot{m} \lesssim 0.1$, the accretion disk is radiatively inefficient accretion flow \citep[see e.g.,][]{Mahadevan1997, XieYuan2012} and could have strong magnetic field owing to the rapid advection. Nevertheless, the thin disk can be formed around $100-1000 R_G$ for a relatively high accretion rate, say $\dot{m} \gtrsim 0.01-0.1$, and in this case, the accumulation of the large-scale magnetic field may be so inefficient that the accretion disk around a BH can be weakly magnetized accretion flow \citep[see e,g.,][]{Esin1997, Kimura2021}. The critical accretion rate above which a MAD is no longer formed is still unclear.

\cite{KayanokiFukazawa2022} reported that GeV-loud objects with high $\dot{m}$ tend to have a low X-ray absorption column density, which implies that a viewing angle may be small. On the other hand, GeV-loud objects with low $\dot{m}$ can have a high column density (see their Figure 5). This implies a large viewing angle, with which emission from the jet should be weaker due to the low Doppler factor. These features may support our conclusions that the low $\dot{m}$ objects emit gamma-rays by MADs, while high $\dot{m}$ objects emit gamma-rays by jets.

The electron heating rate by magnetic reconnection has not been established yet.
We examine another formalism of the electron heating rate given by \cite{Chael2018}. 
The value of the electron heating rate is higher than that of \cite{Hoshino2018}. This results in high optical and X-ray fluxes, which easily overshoot the observational data if we adjust the $\dot{m}$ using gamma-ray data. Thus, more than half of our sample are classified as Bad. This feature is independent of the value of $M$ and $\dot{m}$. 
Thus, the electron heating rate has a strong influence on whether we can explain the GeV gamma-ray data by the MAD model. 

We also apply the MAD model to Sgr A*. Since Sgr A* has a low $\dot{m}$, the gamma-ray emission efficiency is very low, and thus, we cannot explain the gamma-ray data by the MAD model. We conclude that the sources of GeV-TeV gamma-rays are other objects in the Galactic Center. We also estimate the CR intensity of Sgr A* and compare the observed one. Because of low $\dot{m}$, the contribution by Sgr A* with the current activity is too low. The Sgr A* may have been active in the past, and it may contribute to super-knee cosmic rays observed on Earth.
\\ \\
 We thank Masaomi Tanaka for his helpful comments. This work is partly supported by JSPS KAKENHI No. 22K14028 (S.S.K.) and 18H01245 (K.T.). This work is also supported by JST, the establishment of university fellowships towards the creation of science technology innovation, Grant Number JPMJFS2102 (R.K.). S.S.K. acknowledges the support by the Tohoku Initiative for Fostering Global Researchers for Interdisciplinary Sciences (TI-FRIS) of MEXT’s Strategic Professional Development Program for Young Researchers.

\appendix
\section{The critical mass accretion rate for thermal electrons }\label{temax}
The temperature of the thermal particles is obtained by the balance between the heating and the energy loss rates: $Q_i = \Lambda_{{\rm{adv}},i} + \Lambda_{{\rm{rad}},i}$, where $i$ is the particle species, $\Lambda_{{\rm{adv}},i}\approx n_i k_B T_i/t_{\rm{fall}}$ is the advection rate and $\Lambda_{{\rm{rad}},i}\approx n_i k_B T_i/t_{{\rm{rad}},i}$ is the radiation cooling rate. The thermal protons do not cool in the range of our interest, and thus, $\Lambda_{{\rm rad},p}=0$. Then, the proton temperature is always given by $\Lambda_{{\rm adv},p}=Q_p$. For a very low accretion rate, advection is dominant even for thermal electrons. This leads to $\Lambda_{\rm adv,e}=Q_e$, and then, we obtain $T_e/T_p=Q_e/Q_p$. For the range of our interest, the thermal synchrotron is the most efficient cooling process, whose cooling rate is given in \cite{KimuraMurase2021}. Equating the thermal synchrotron cooling rate to the heating rate with the elecrtron temperature determined by advection, we obtain the critical mass accretion rate above which the radiative cooling is effective:
\begin{equation}
\dot{m} \lesssim \dot{m}_{\rm{crit}} = 1.5\times 10^{-8} \ M_9 \mathcal{R}_1^{35/2} \alpha_{-0.5}^3 \beta_{-1}^3 \left(\frac{\epsilon_{\rm{dis}}(1-\epsilon_{\rm{NT}})}{0.15 \cdot 0.67}\right)^2 {x_M}_3^{-6},
\end{equation}
where $x_M=\nu_{\rm thml,peak}/\nu_{\rm syn}$, $\nu_{\rm{thml, peak}}$ is the peak frequency of the synchrotron spectrum by the thermal electrons, $\nu_{\rm syn}=3\theta_e^2eB/(4\pi m_ec)$ is the synchrotron frequency, and $\theta_e=k_BT_e/(m_ec^2)$.
$\dot{m}_{\rm crit}$ strongly depends on $\mathcal{R}$ and $x_M$. For our fiducial parameter set, $\dot{m}_{\rm crit}$ is very low, and the radiation cooling is dominant for all of our radio-galaxy samples. On the other hand, advection is dominant for the cases with $\mathcal{R}=30$, i.e., for the Good and Bad objects. Also, we find that advection is dominant for Sgr A*, due to a small value of $x_M\simeq 54$.

The radiation timescale is shorter than the advection timescale for the thermal electrons if $\dot{m}>\dot{m}_{\rm crit}$. The radiation cooling leads to a lower electron temperature than that determined by advection. Thus, the electron temperature should be in the range of  
\begin{equation}
T_e \leq \frac{Q_e}{Q_p} T_p .
\end{equation}

\section{Photon spectra and resulting quantities for various radio galaxies} \label{ap_hoshino}
\begin{figure}[tb]
 \gridline{\fig{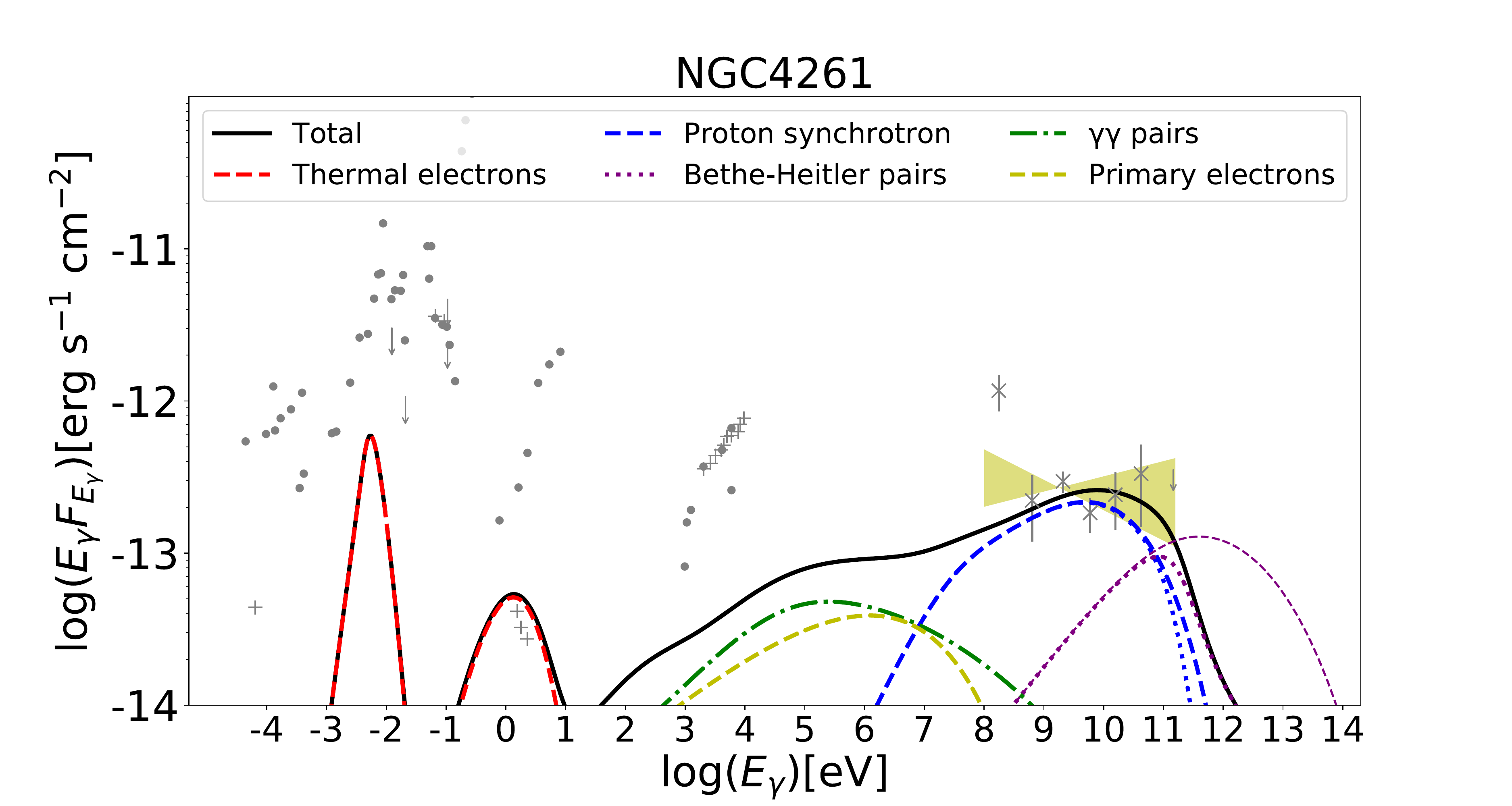}{0.33\textwidth}{}
          \fig{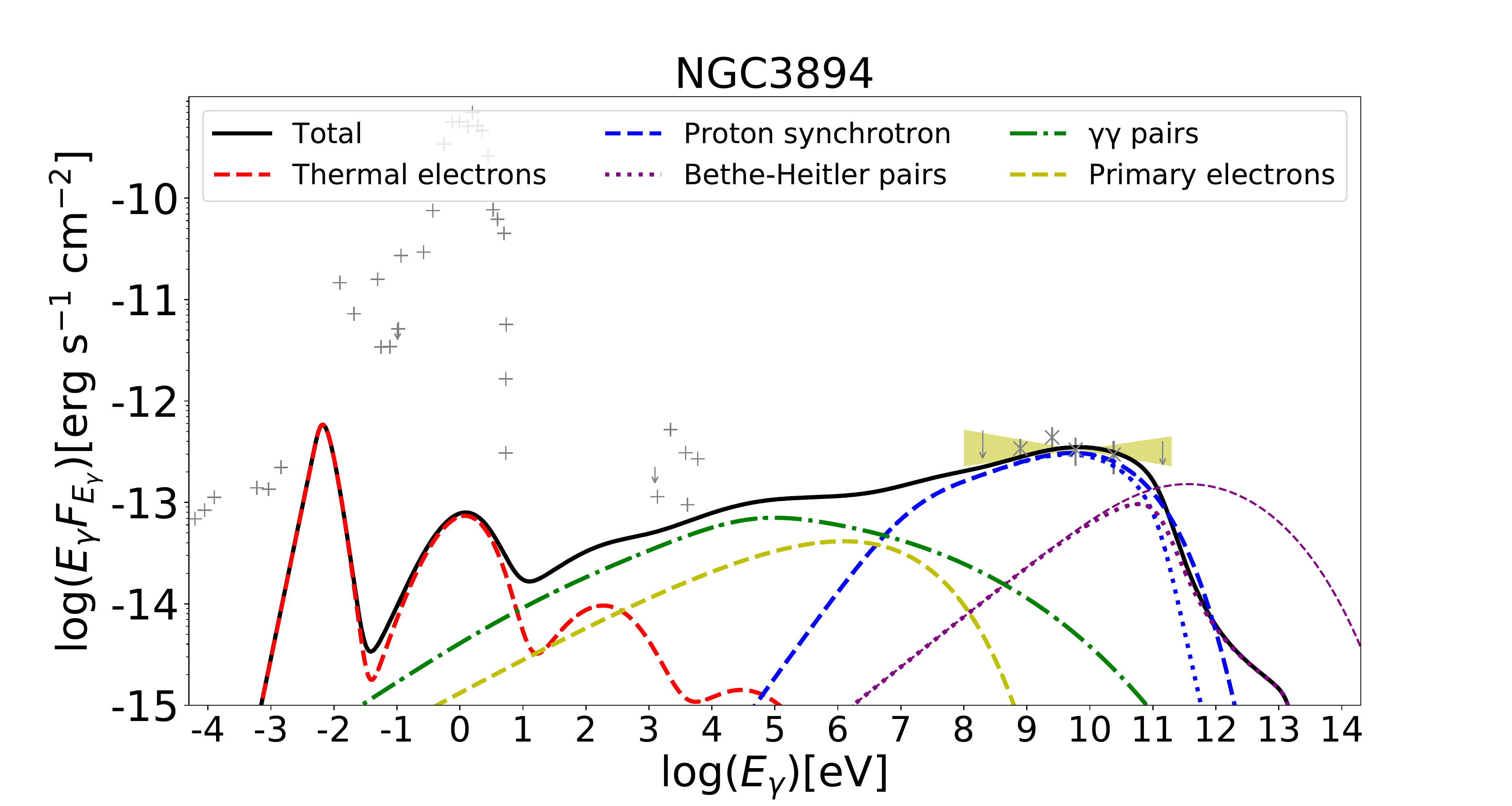}{0.33\textwidth}{}
          \fig{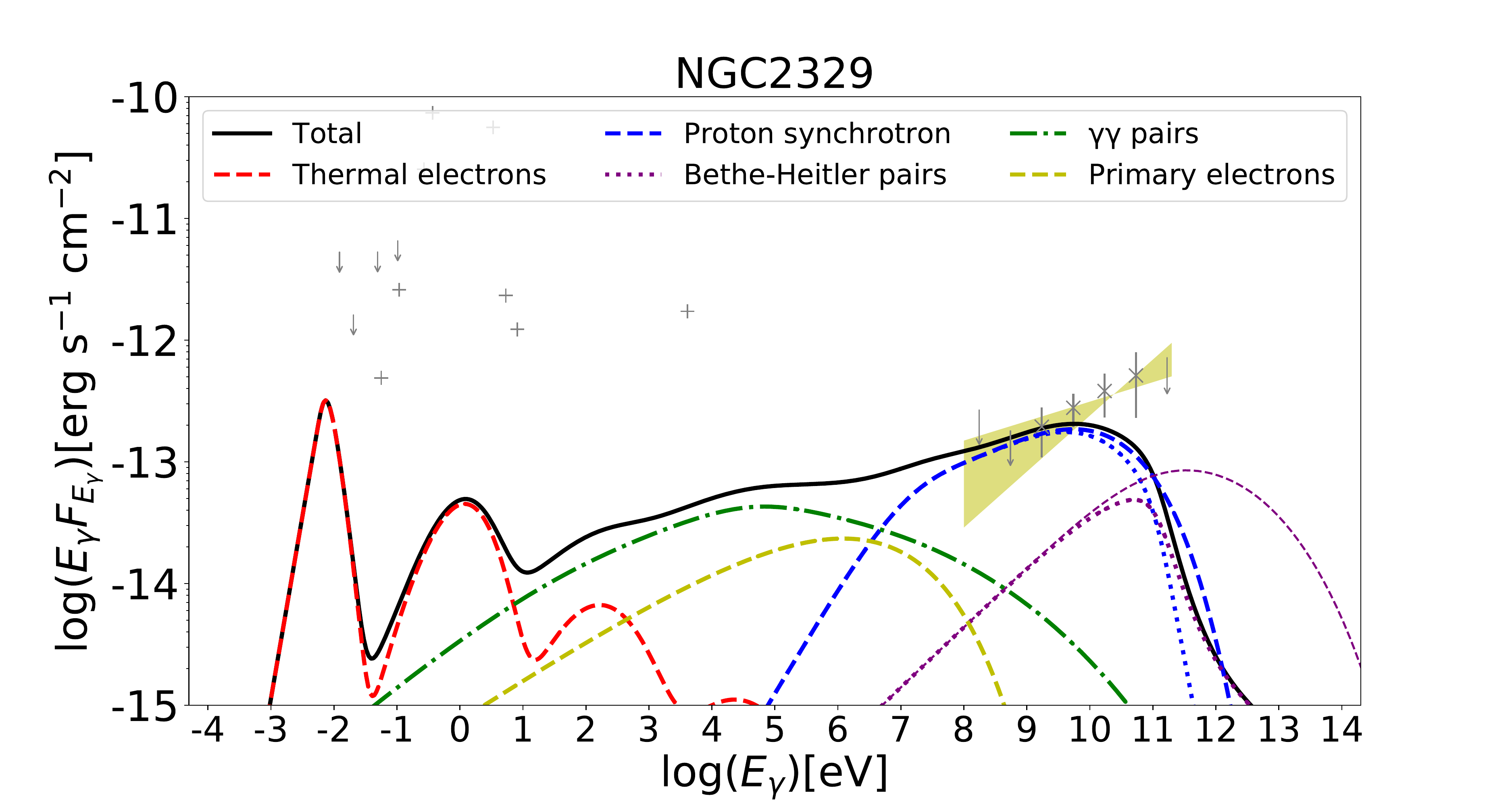}{0.33\textwidth}{}
          }
\gridline{\fig{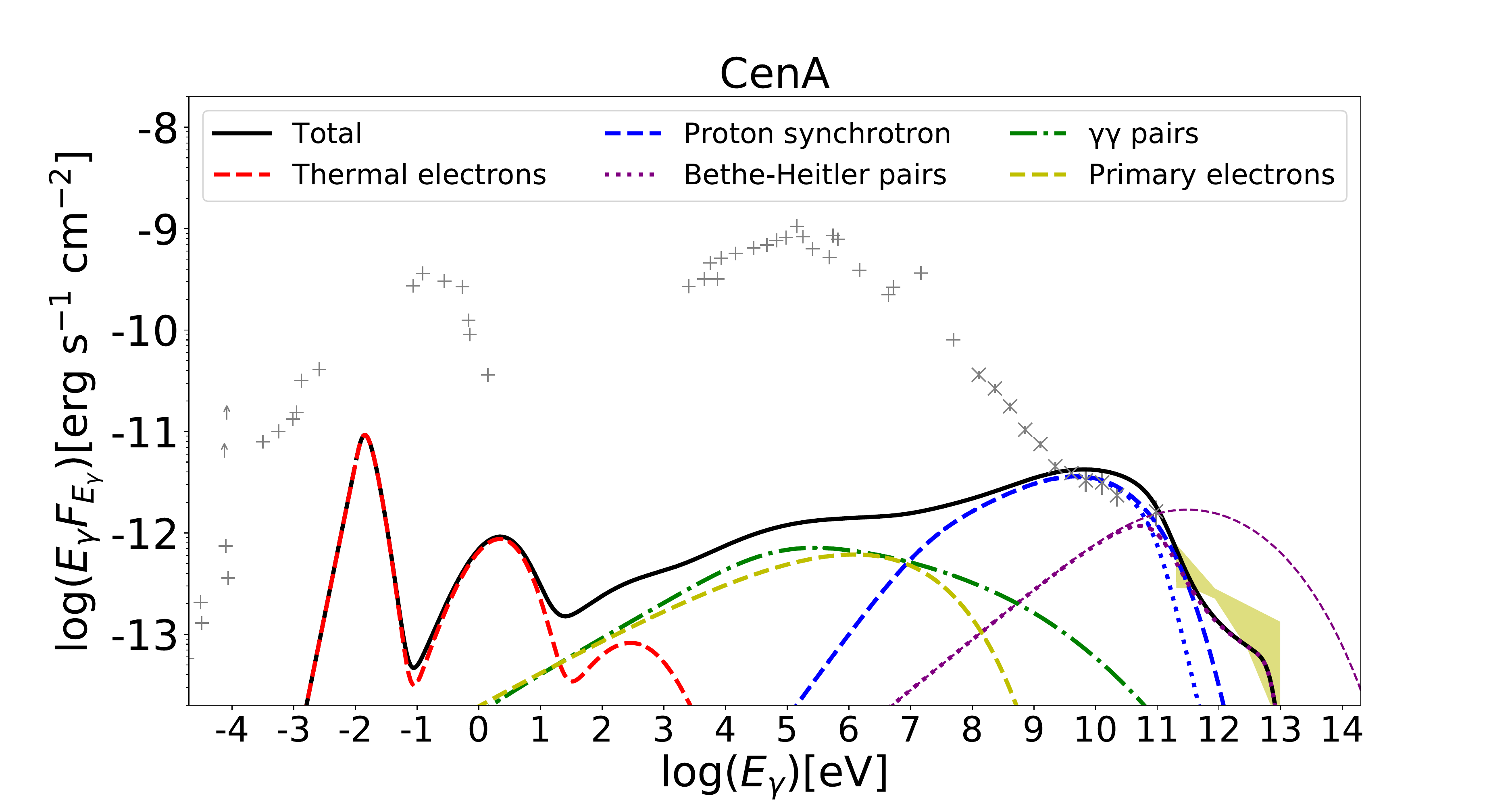}{0.33\textwidth}{}
          }
 \caption{Photon spectra for the Excellent objects. The line types are the same as Figure \ref{fig_typ}. Data points are taken from \cite{Tomar2021} for NGC 4261, \cite{Principe2020} for NGC 3894, \cite{Rulten2020} for NGC 2329, and \cite{HESS2020} and \cite{Abdo2010} for Cen A. Other data points are taken from NED ({\url{http://ned.ipac.caltech.edu}}).}
 \label{fig_vg}
\end{figure}

\begin{figure}[tb]
 \gridline{\fig{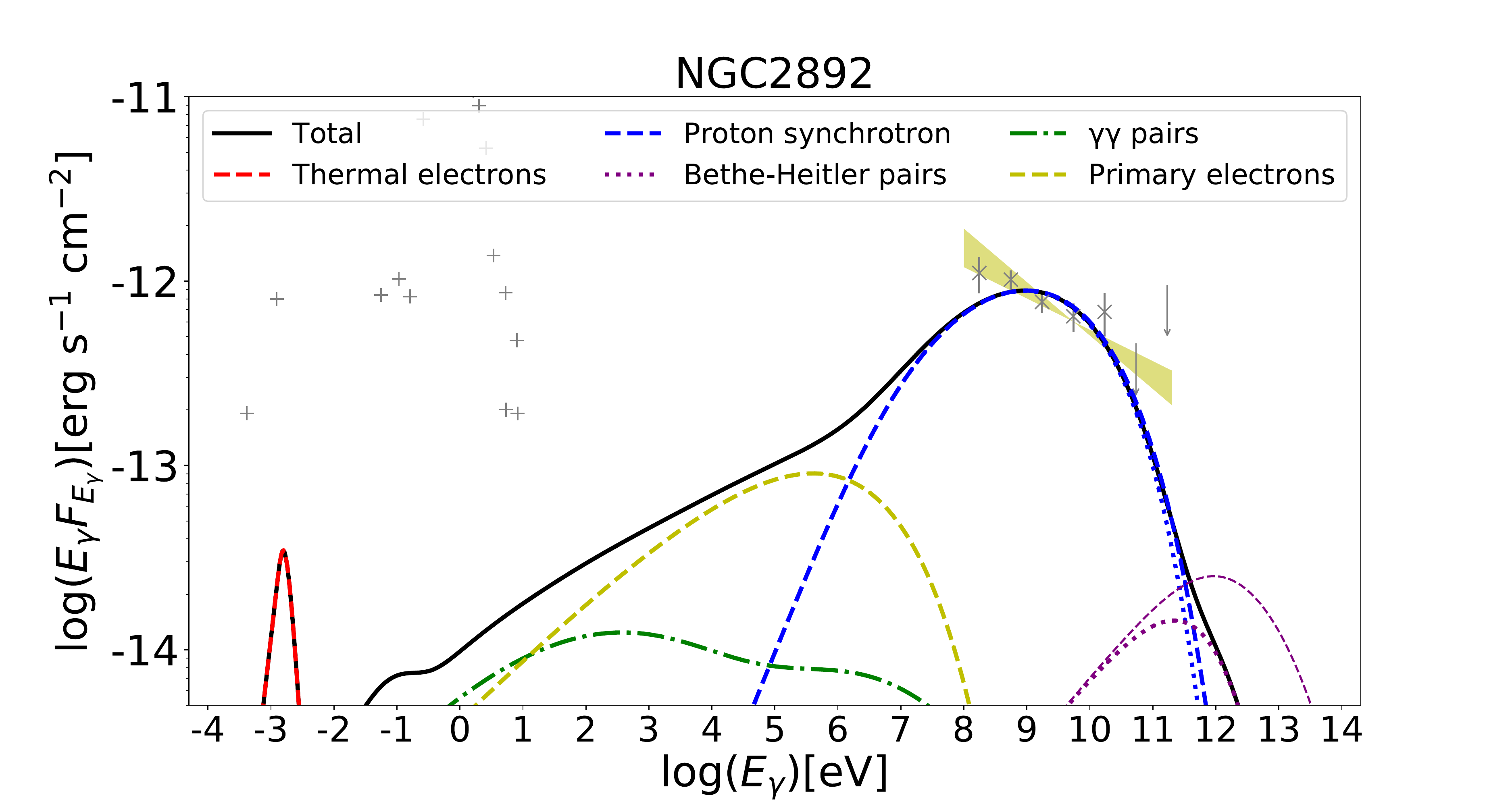}{0.33\textwidth}{}
           \fig{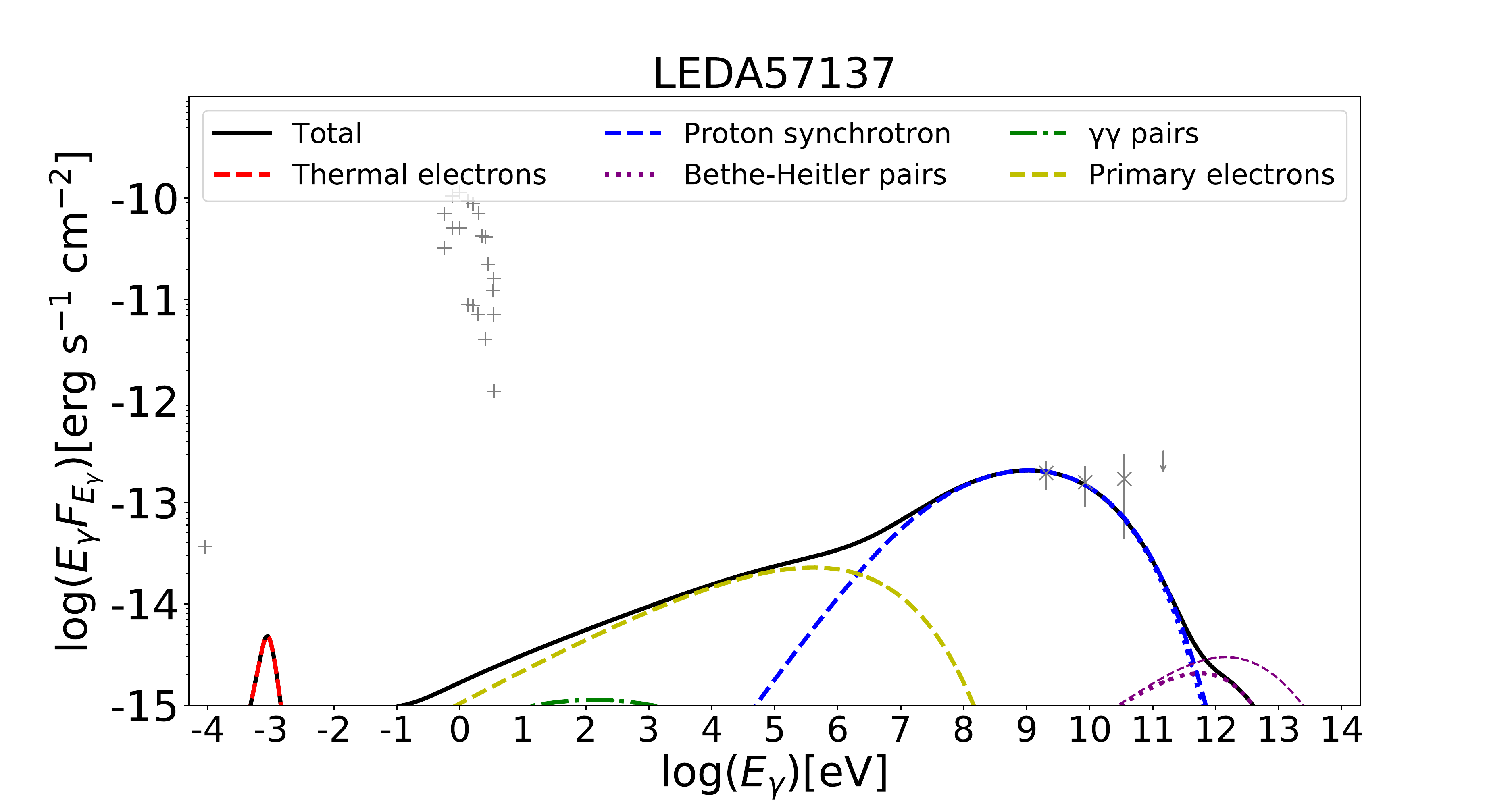}{0.33\textwidth}{}
           \fig{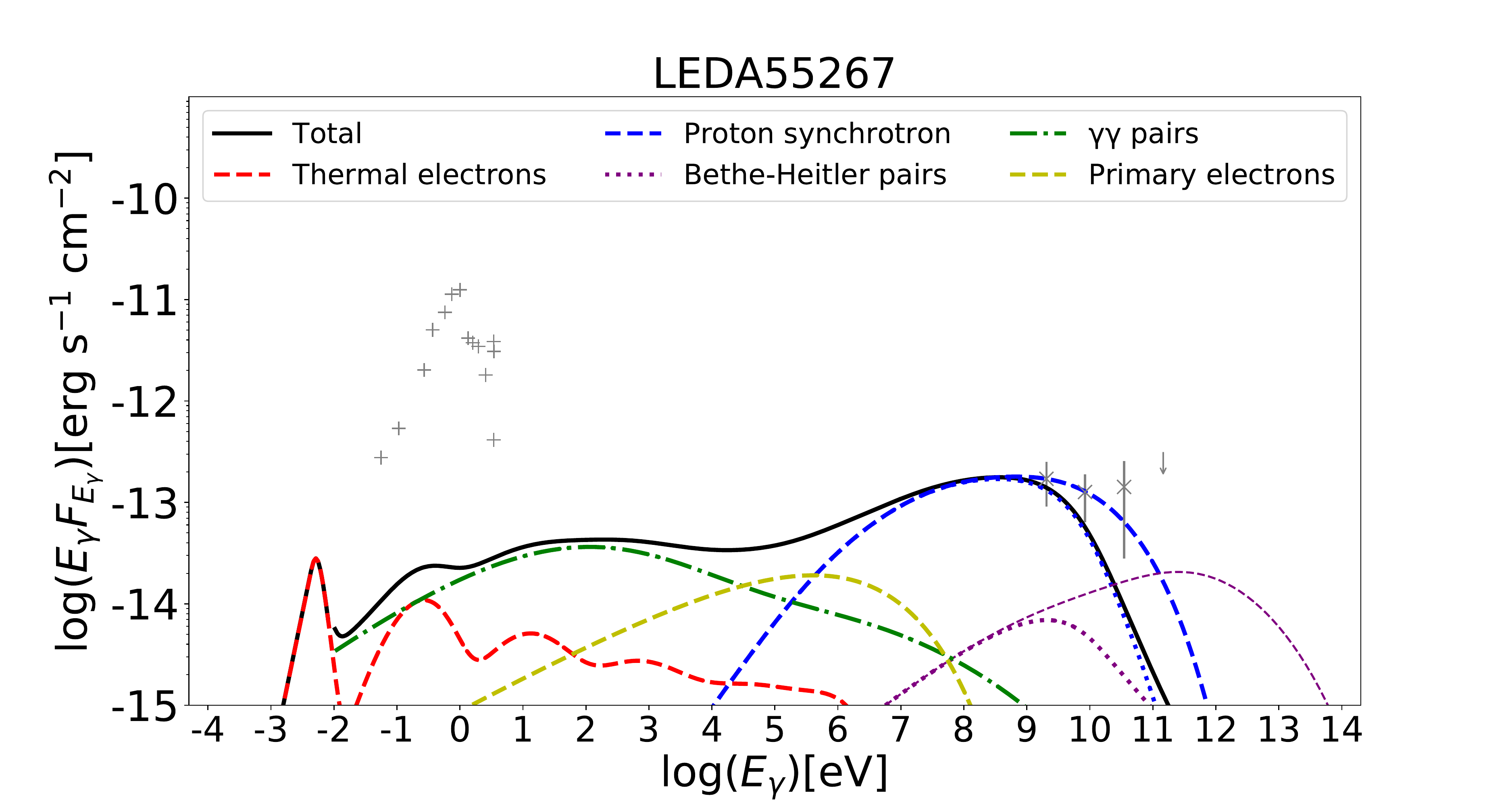}{0.33\textwidth}{}
          }
\gridline{\fig{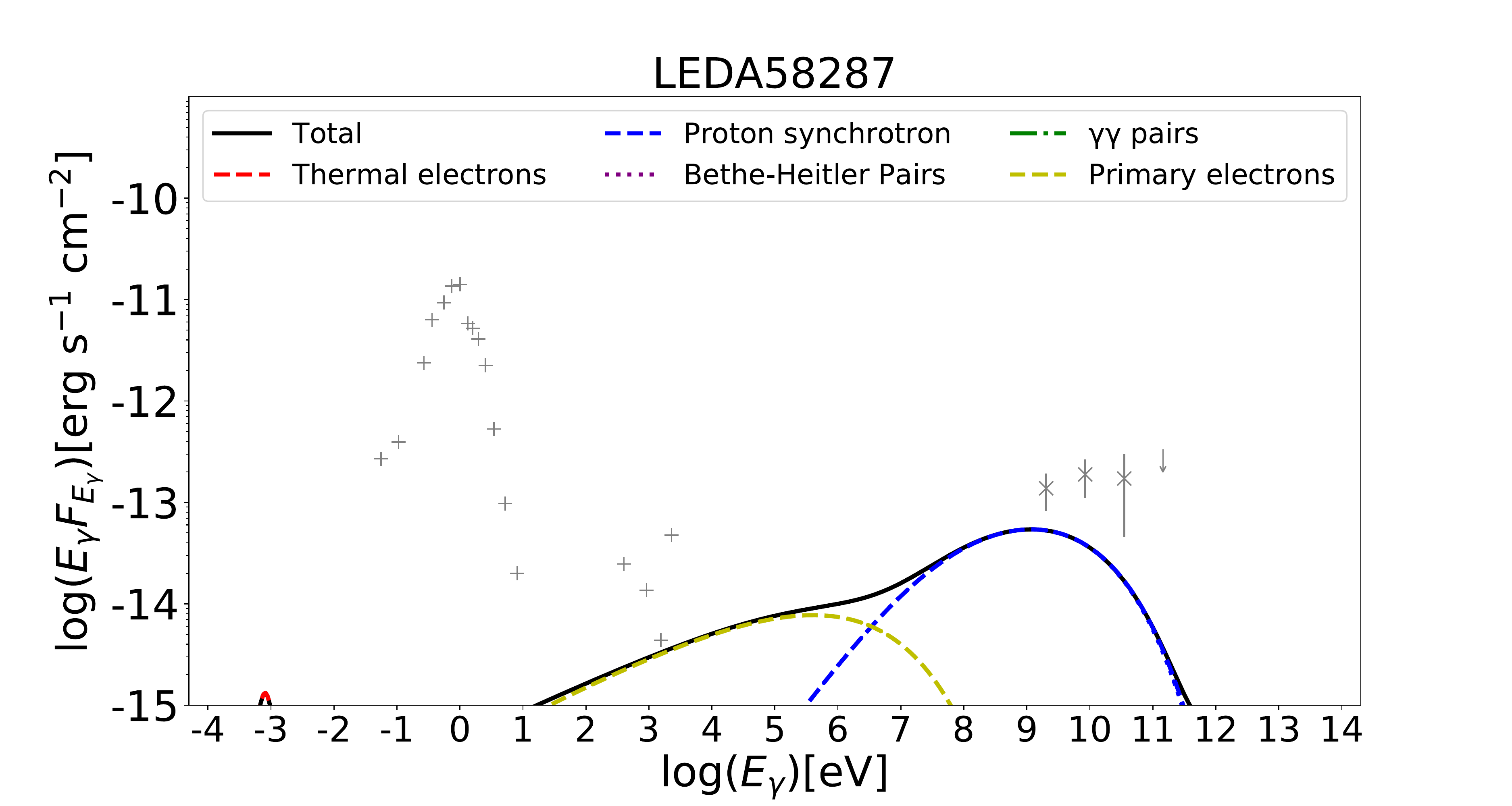}{0.33\textwidth}{}
            \fig{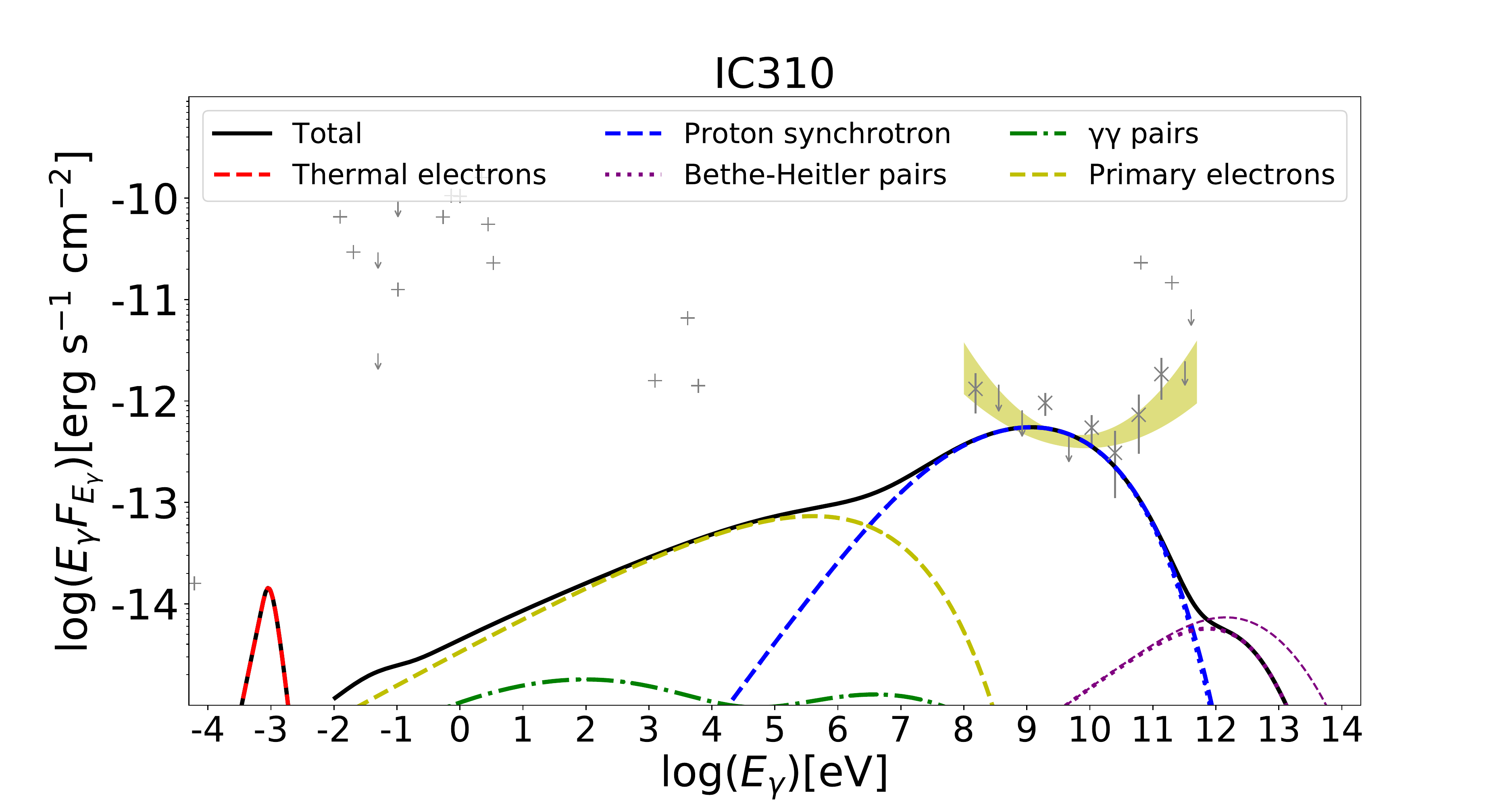}{0.33\textwidth}{}
          }
 \caption{ Photon spectra for the Good objects. The line types are the same as Figure \ref{fig_typ}. Data points in GeV gamma-rays are taken from \cite{Tomar2021} for NGC 2892, \cite{Paliya2021} for LEDA 57137, LEDA 55267, and LEDA 58287, and \cite{Graham2019} for IC 310. Other data points are taken from NED.}
 \label{fig_gd}
\end{figure}

\begin{figure}[tb]
 \gridline{\fig{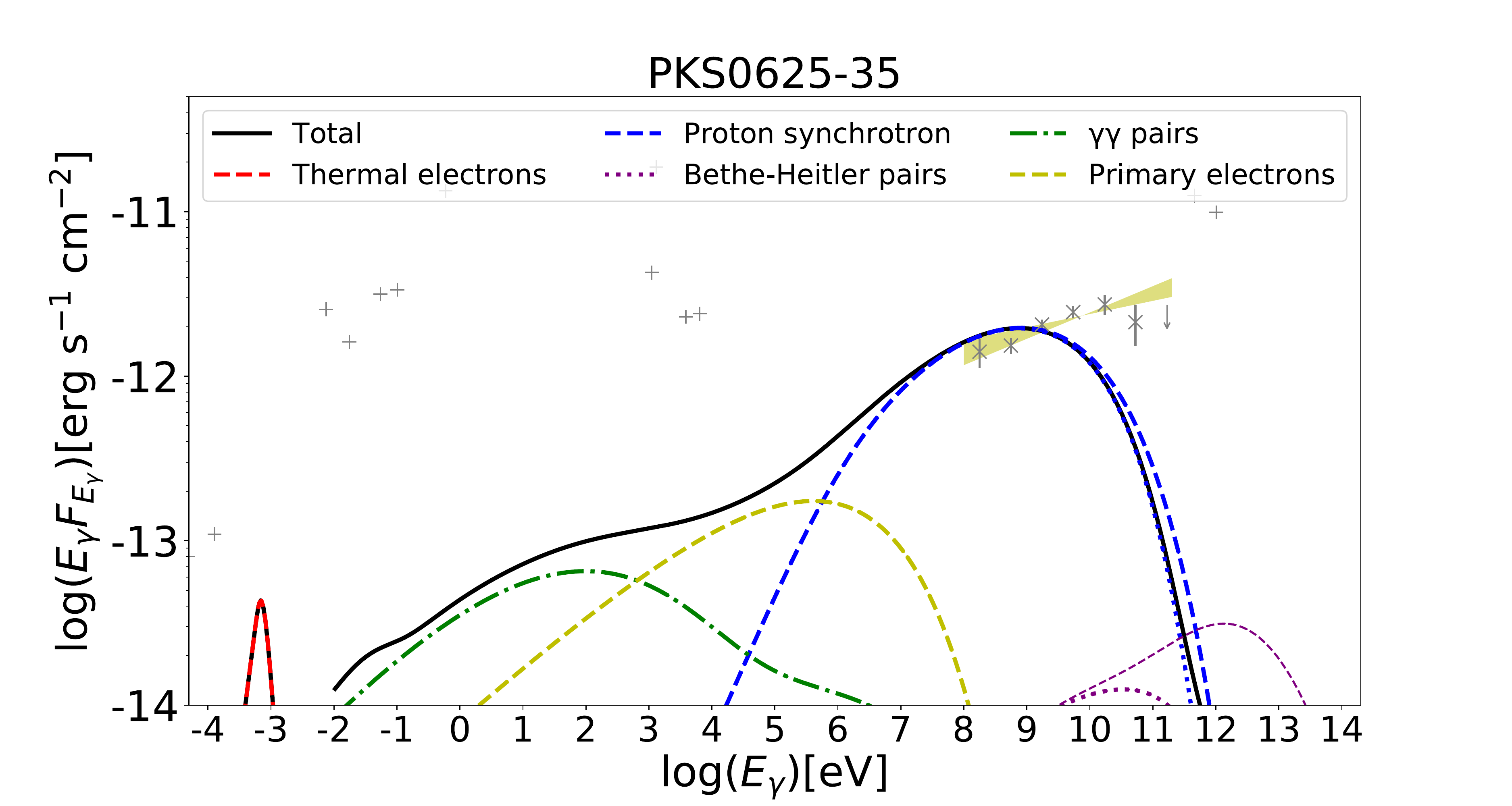}{0.33\textwidth}{}
          \fig{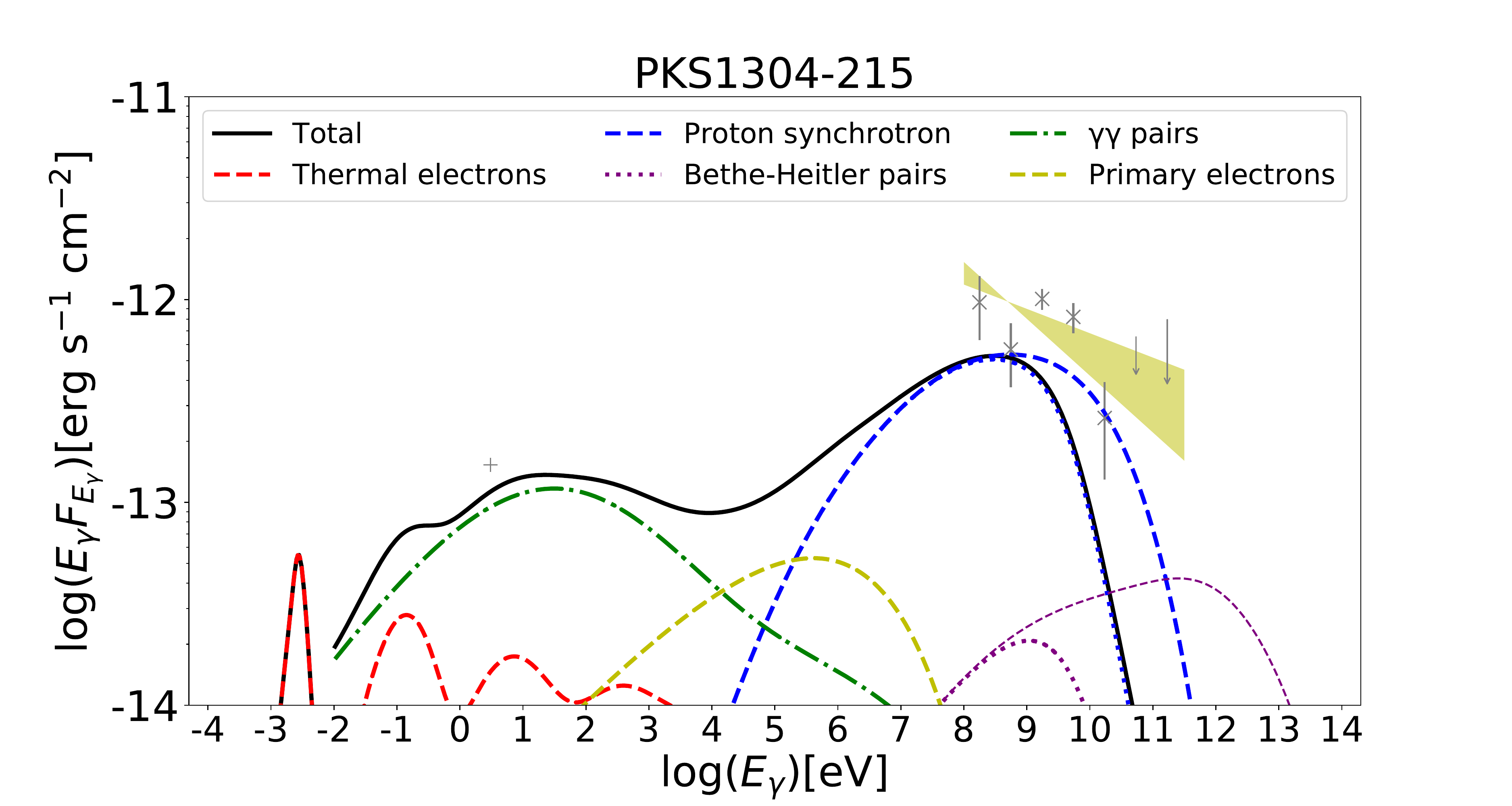}{0.33\textwidth}{}
          \fig{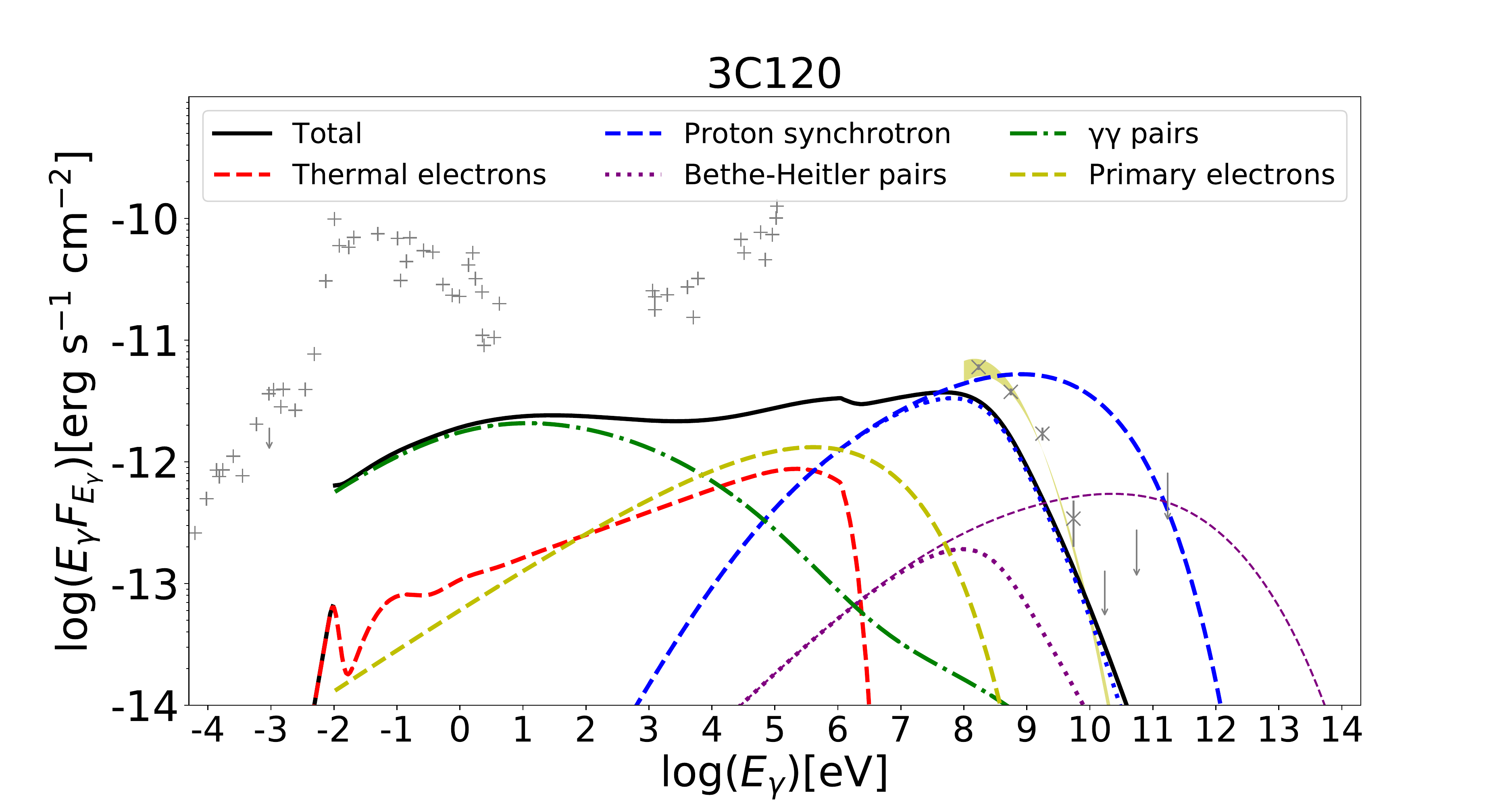}{0.33\textwidth}{}
          }
\gridline{\fig{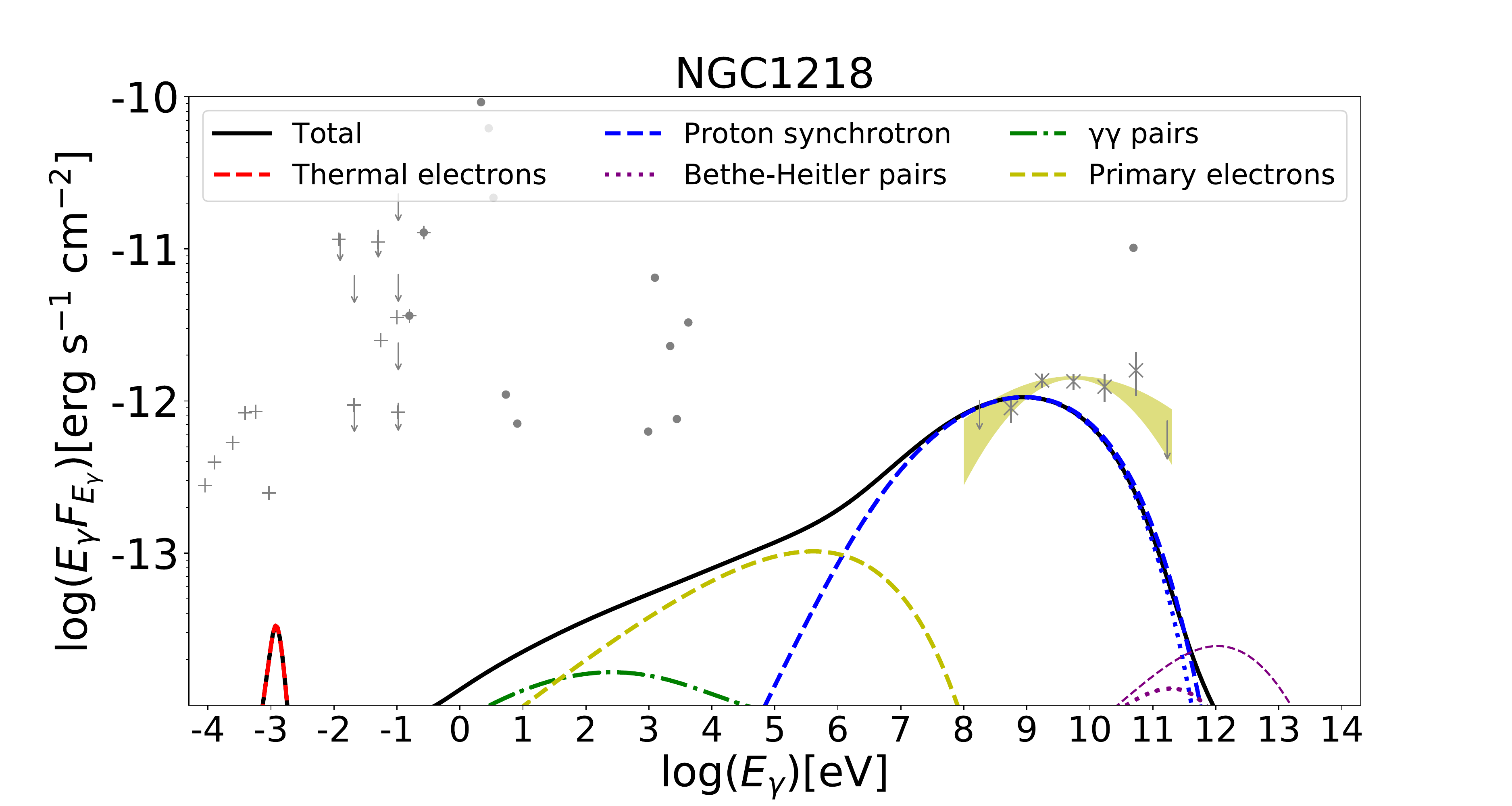}{0.33\textwidth}{}
            \fig{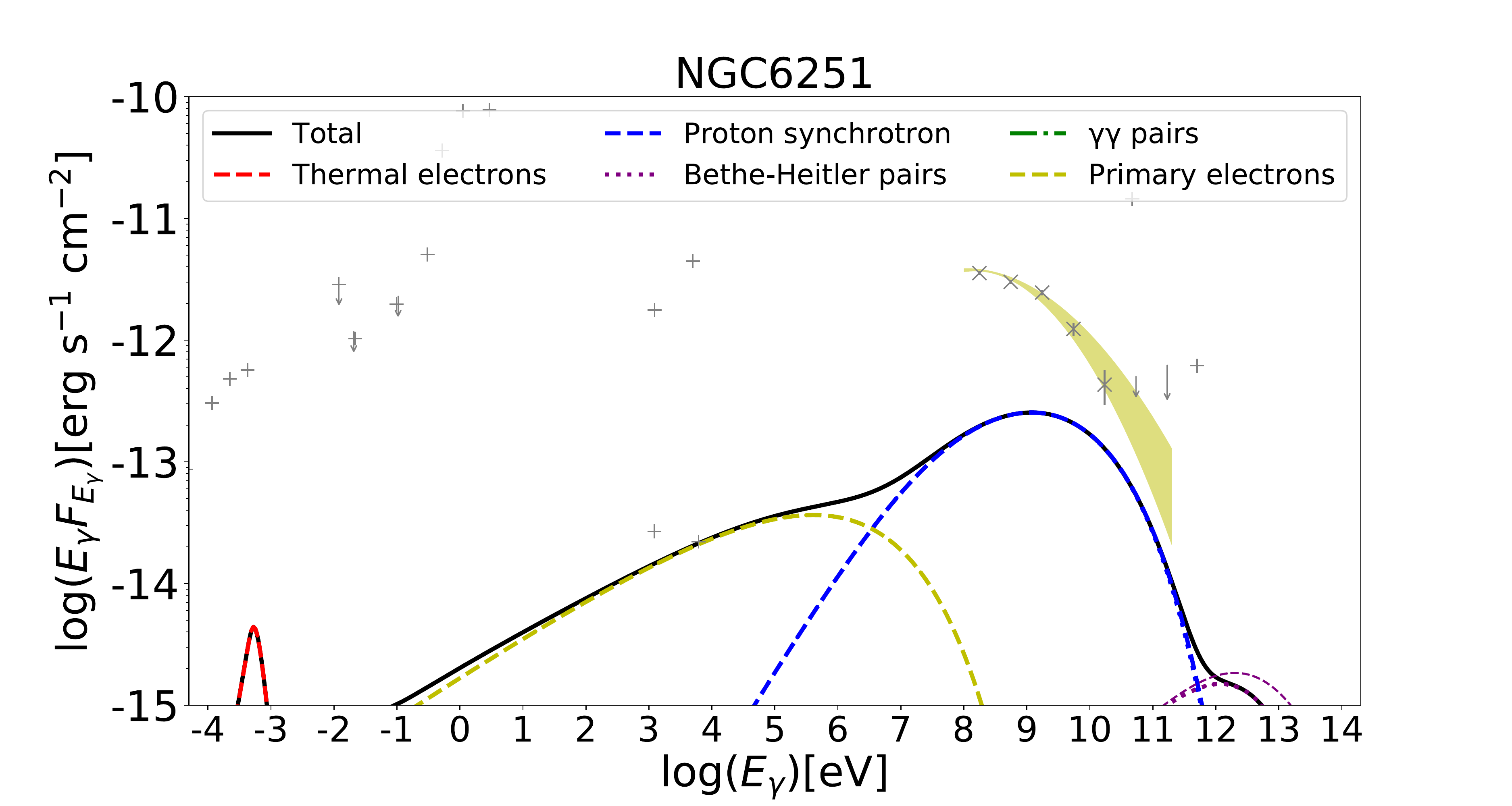}{0.33\textwidth}{}
            \fig{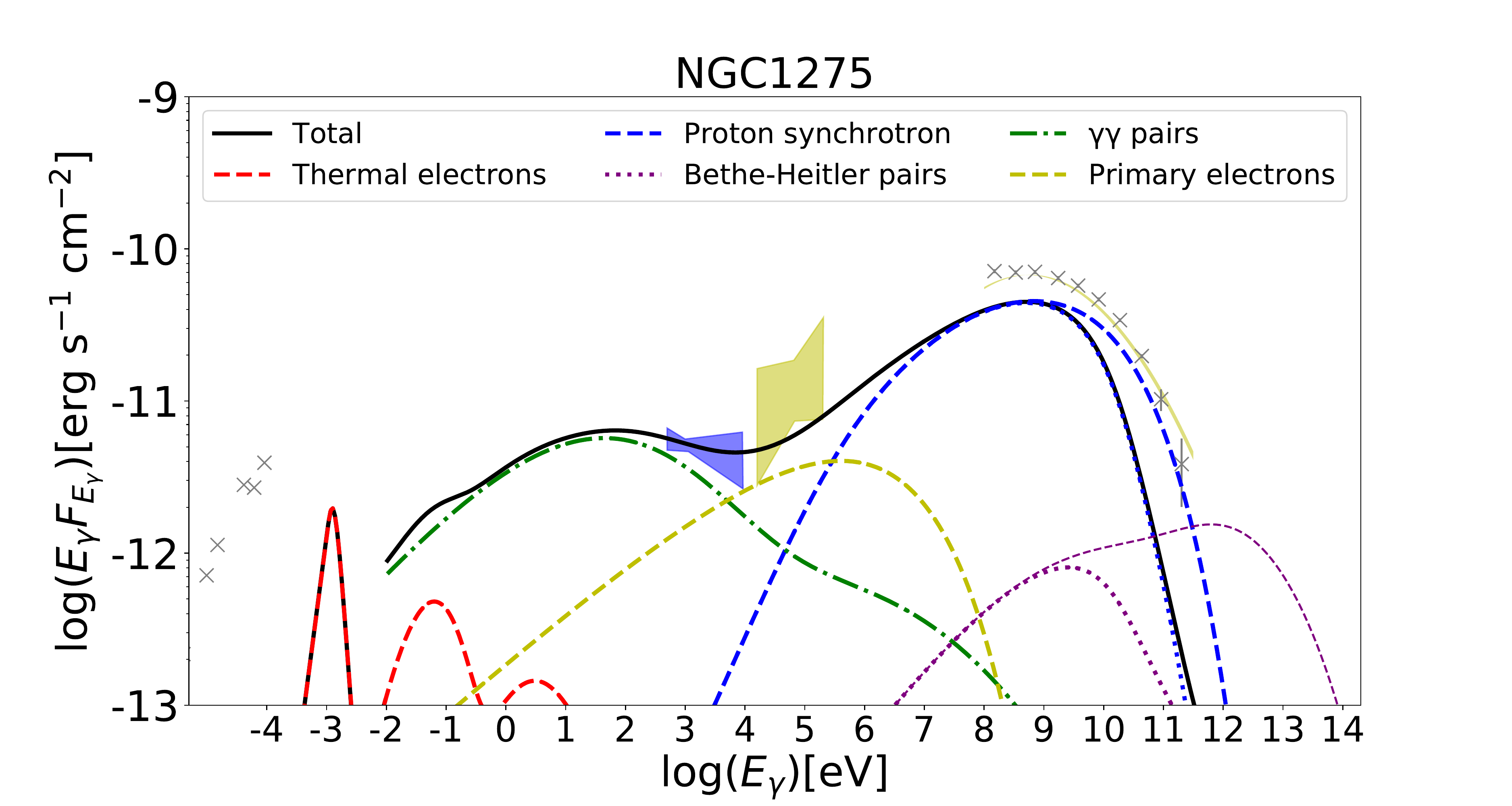}{0.33\textwidth}{}
          }

 \caption{Photon spectra for the Bad objects. The line types are the same as Figure \ref{fig_typ}. Data points are taken from \cite{Rulten2020} for PKS 0625-35, PKS 1304-215, 3C 120, NGC 1218, and NGC 6251, and \cite{Tomar2021} for NGC 1275. Other data points are taken from NED.}
 \label{fig_bd}
\end{figure}

We show the photon spectra for the various radio galaxies with the electron heating rate of \cite{Hoshino2018}. The photon spectra for the Excellent, Good, and Bad objects are shown in Figure \ref{fig_vg}, Figure \ref{fig_gd}, and Figure \ref{fig_bd}, respectively. We tabulate the resulting quantities for the Excellent, Good, and Bad objects in Table \ref{tb_vg}, Table \ref{tb_gd}, and Table \ref{tb_bd}, respectively. For LEDA 58287, the MAD model underpredicts the GeV gamma-ray flux. However, because of their large error bars and the small number of data points, this object statistically results in a Good object.
\begin{deluxetable}{ccccccc}
\tablenum{3}
\tablecaption{Quantities for the Excellent objects.
\label{tb_vg}}
\tablewidth{0pt}
\tablehead{
\colhead{Name} & \colhead{Mass $[M_\odot]$}  & \multicolumn{0}{c}{Distance [Mpc]} & \colhead{$\dot{m}$} & \colhead{$\chi^2/\nu$} & \colhead{$Q$}
}
\startdata
    NGC 4261 & $4.9 \times 10^{8}$ & 35.1 & $2.2 \times 10^{-4}$ & 1.1  & 0.35 \\
    NGC 3894 & $5.4 \times 10^{8}$ & 50.1 & $4.5 \times 10^{-4}$ & 0.47 & 0.70  \\
    NGC 2329 & $4.9 \times 10^{8}$ & 71.1 & $5.7 \times 10^{-4}$ & 1.4  & 0.24 \\ 
    Cen A   & $5.5 \times 10^{7}$ & 3.8 & $3.9 \times 10^{-4}$ & 3.3 & 0.011 \\
\enddata
\tablecomments{The references for BH masses and distances are \cite{Tomar2021} for NGC 4261, \cite{Mould2012}, \cite{Balasubramaniam2021} for NGC 3894, \cite{Das2021}, \cite{Ellis2006} for NGC 2329, and \cite{Harris2010} and \cite{EHT2021CenAjet} for Cen A.}
\end{deluxetable}

\begin{deluxetable}{ccccccc}
\tablenum{4}
\tablecaption{Same as Table \ref{tb_vg}, but for Good objects. The BH mass are enhanced by a factor of 3 from the values in the references. \label{tb_gd}}
\tablewidth{0pt}
\tablehead{
\colhead{Name} & \colhead{Mass $[M_\odot]$}  & \multicolumn{0}{c}{Distance [Mpc]} & \colhead{$\dot{m}$} & \colhead{$\chi^2/\nu$} & \colhead{$Q$}
}
\startdata
    NGC 2892 & $8.4 \times 10^{8}$ & 86.2 & $1.7 \times 10^{-3}$ & 1.53 & 0.19 \\
    LEDA 57137 & $1.5\times 10^{9}$ & 171 & $9.2 \times 10^{-4}$ & 0.34 & 0.71 \\
    LEDA 55267 & $4.8 \times 10^{8}$ & 327 & $8.9\times10^{-3}$ & 1.36 & 0.26 \\
    LEDA 58287 & $9.6\times 10^{8}$ & 185 & $5.7\times10^{-4}$ & 3.82  & 0.02 \\
    IC 310 & $9.0 \times 10^{8}$ & 63 & $6.7\times10^{-4}$ & 2.74 & 0.018 \\
\enddata
\tablecomments{The references for BH masses, distances are \cite{Beifiori2012} for NGC 2892,  \cite{Paliya2021} and NED for LEDA 55267, LEDA 57137, and LEDA 58287, and \cite{Schulz2015} and \cite{Ellis2006} for IC 310.}
\end{deluxetable}

\begin{deluxetable}{ccccccc}
\tablenum{5}
\tablecaption{Same as Table \ref{tb_vg}, but for Bad objects. The BH mass are enhanced by a factor of 3 from the values in the references. \label{tb_bd}}
\tablewidth{0pt}
\tablehead{
\colhead{Name} & \colhead{Mass $[M_\odot]$}  & \multicolumn{0}{c}{Distance [Mpc]} & \colhead{$\dot{m}$} & \colhead{$\chi^2/\nu$} & \colhead{$Q$}
}
\startdata
    PKS 0625-35 & $9 \times 10^{9}$ & 243.7 & $2.4 \times 10^{-3}$ & 12 & $1.0 \times 10^{-11}$ \\
    PKS 1304-215 & $3\times 10^{9}$ & 564 & $1.1 \times 10^{-2}$ & 13 & $2.5\times 10^{-10} $ \\
    3C 120 & $1.89 \times 10^{8}$ & 139 & $1.0 \times 10^{-1}$ & 65 & $3.0\times10^{-42}$ \\
    NGC 1275 & $7.2\times 10^{9}$ & 70.1 & $5.7 \times 10^{-3}$ & 289 & 0.0  \\ 
    NGC 1218 & $1.65 \times 10^{9}$ & 116 & $1.7 \times 10^{-3}$ & 7.4 & $6.0 \times 10^{-6}$ \\
    NGC 6251 & $1.84 \times 10^{9}$ & 104.6 & $4.5 \times 10^{-4}$ & $4.1 \times 10^{2}$ & 0.0  \\
 \enddata
\tablecomments{The references for BH masses, distances are \cite{Rani2019} and \cite{Sahakyan2018} for PKS 0625-35, NED for PKS 1304-215, \cite{Hlabahte2020} for 3C 120, \cite{Scharwachter2013} and \cite{Tomar2021} for NGC 1275, \cite{Falcke2004} for NGC 1218, and \cite{Graham2008} for NGC 6251. For PKS 1304-215, there is no data for BH mass, and we assume the BH mass as $10^9 M_\odot$.}
\end{deluxetable}

\section{Photon spectra and resulting quantities for the various radio galaxies with another electron heating prescription} \label{chael}
\begin{figure}[tb]
\gridline{\fig{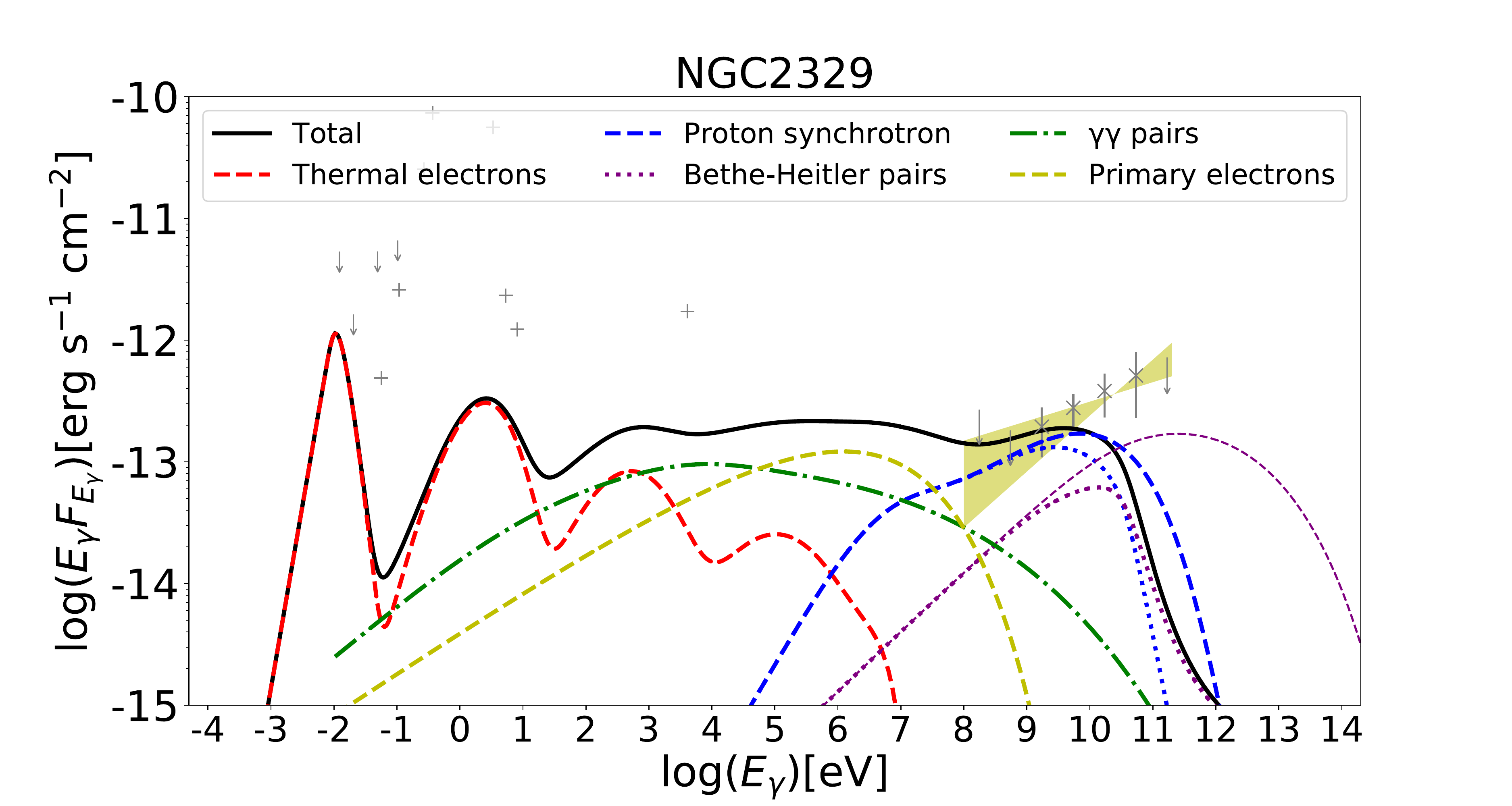}{0.33\textwidth}{}
         \fig{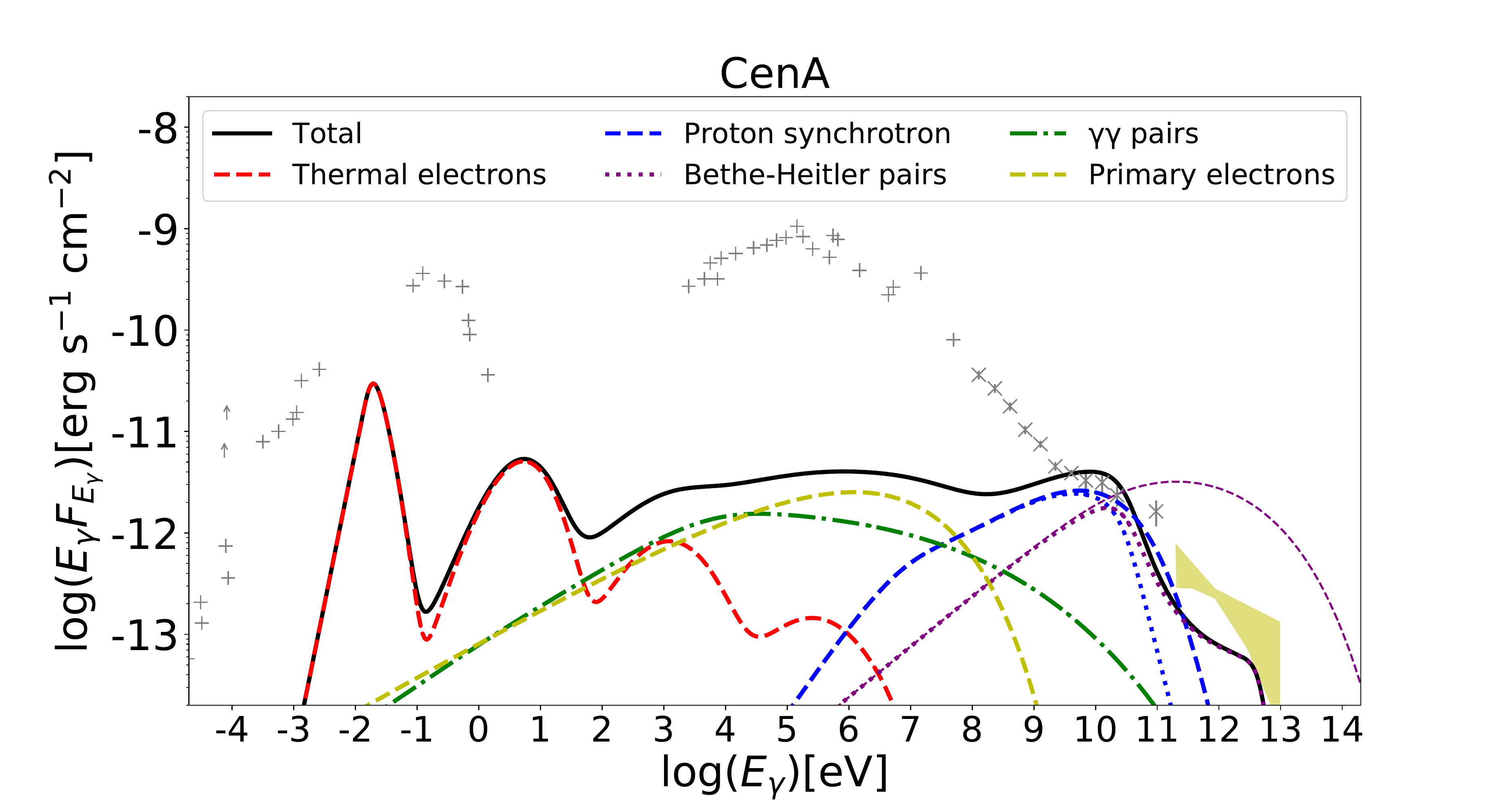}{0.33\textwidth}{}
            }
 \caption{Same as Figure \ref{fig_vg}, but with the electron heating rate given by \cite{Chael2018}.}
 \label{fig_vg_chael}
\end{figure}

\begin{figure}[tb]
\gridline{\fig{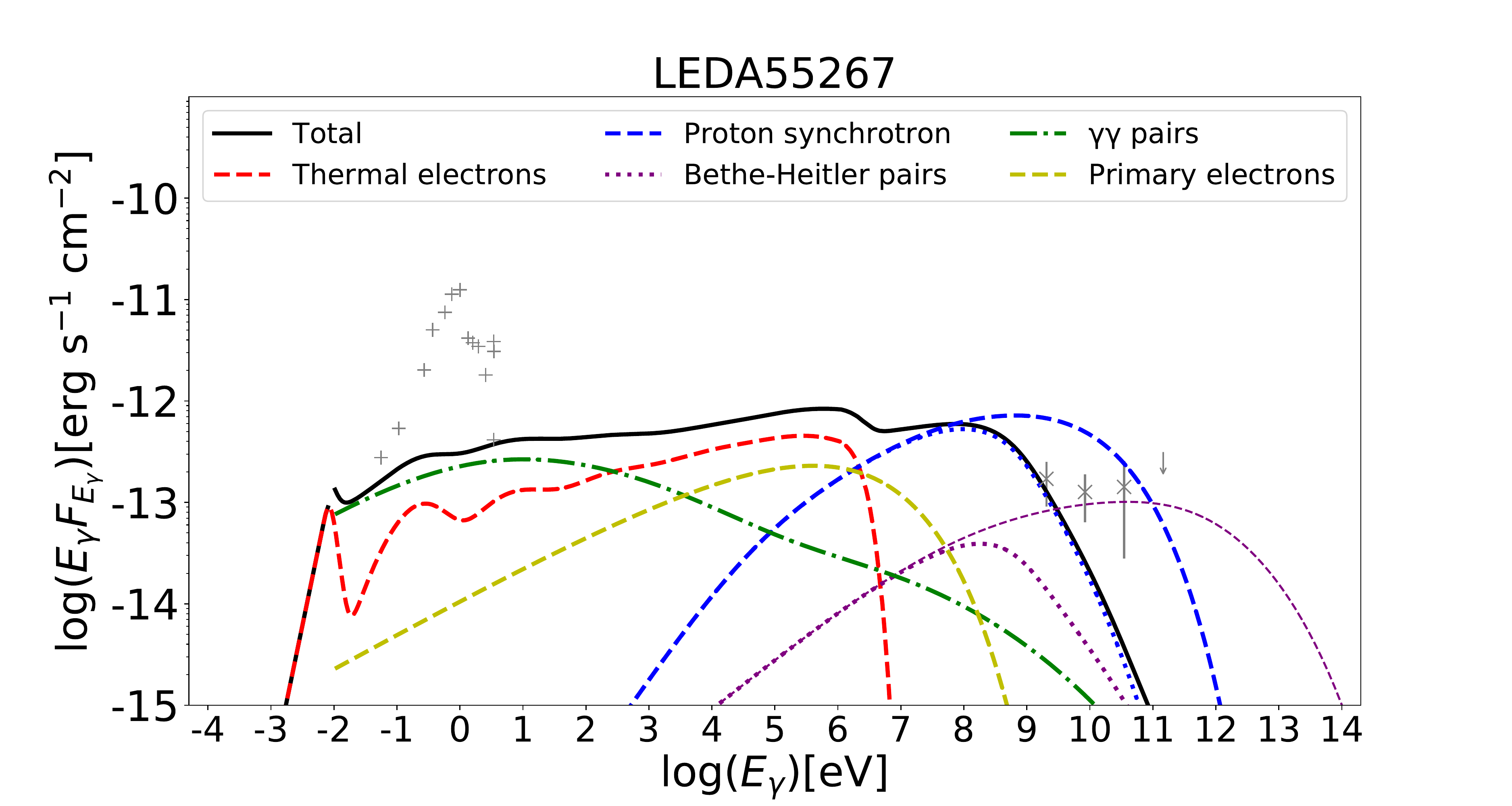}{0.33\textwidth}{}
            \fig{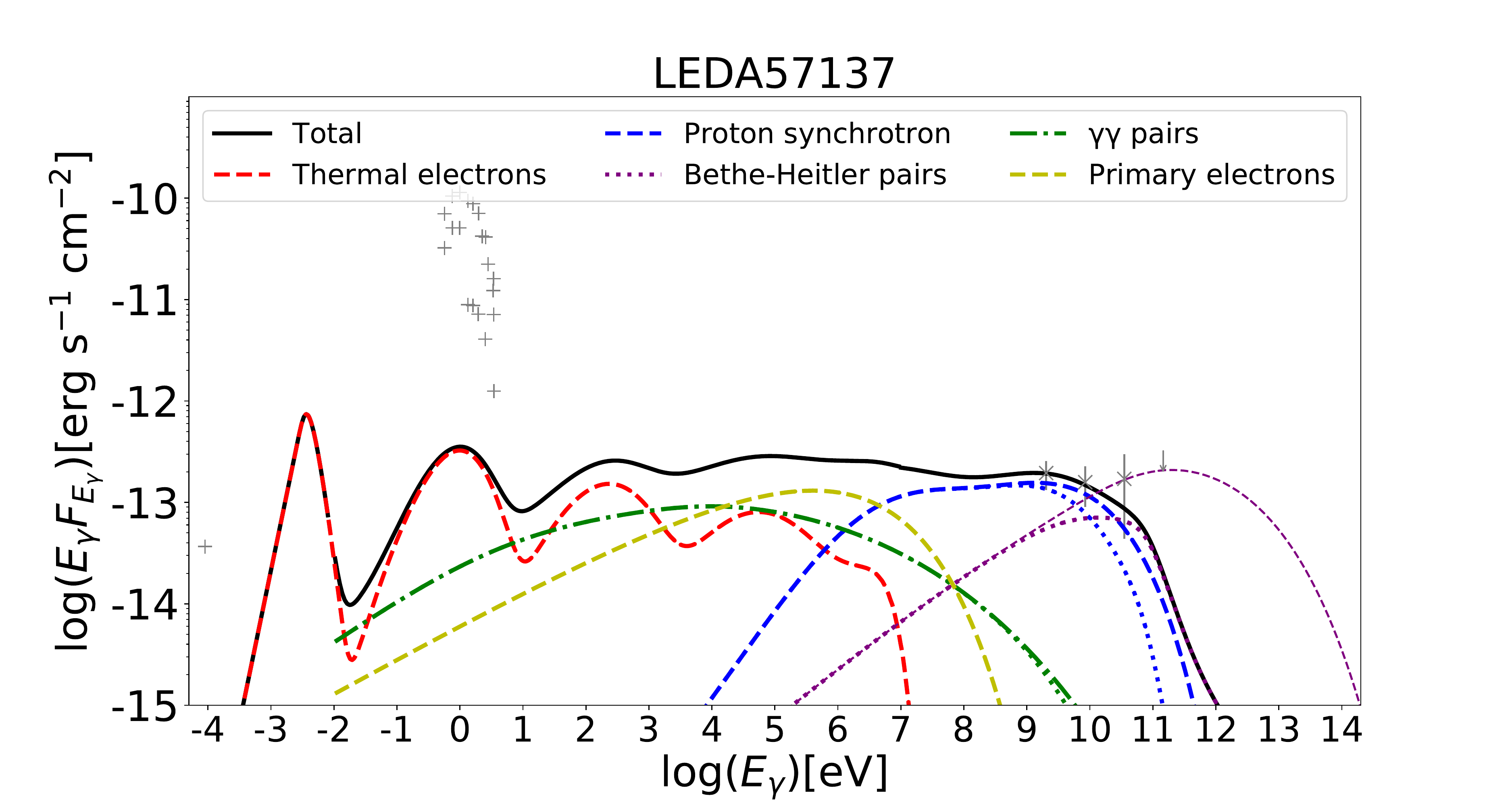}{0.33\textwidth}{}
            \fig{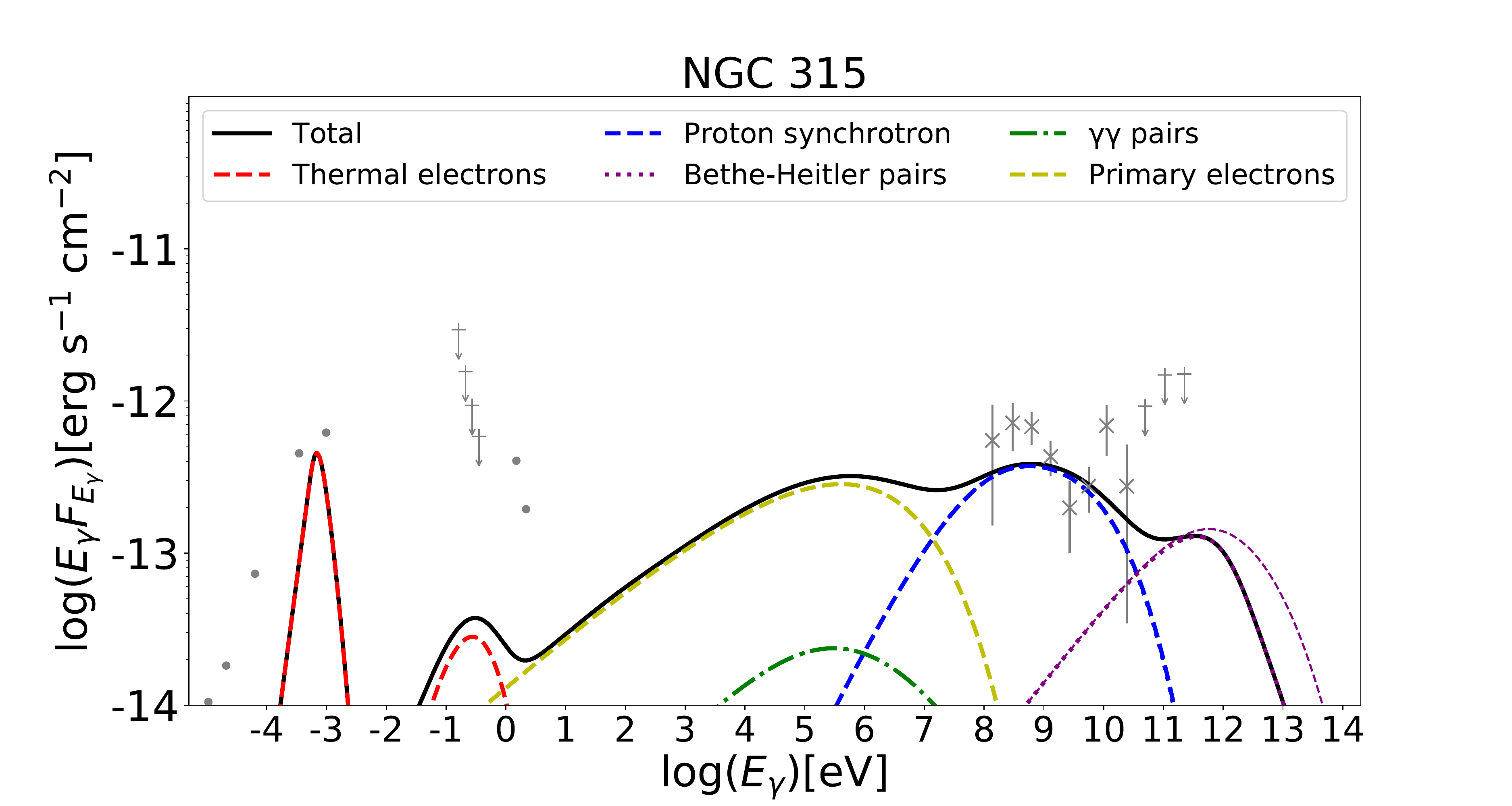}{0.33\textwidth}{}
          }
 \caption{Same as Figure \ref{fig_gd}, but with the electron heating rate given by \cite{Chael2018}. Data points are taken from \cite{Menezes2020} for NGC 315.}
 \label{fig_gd_chael}
\end{figure}

\begin{figure}[tb]
\gridline{\fig{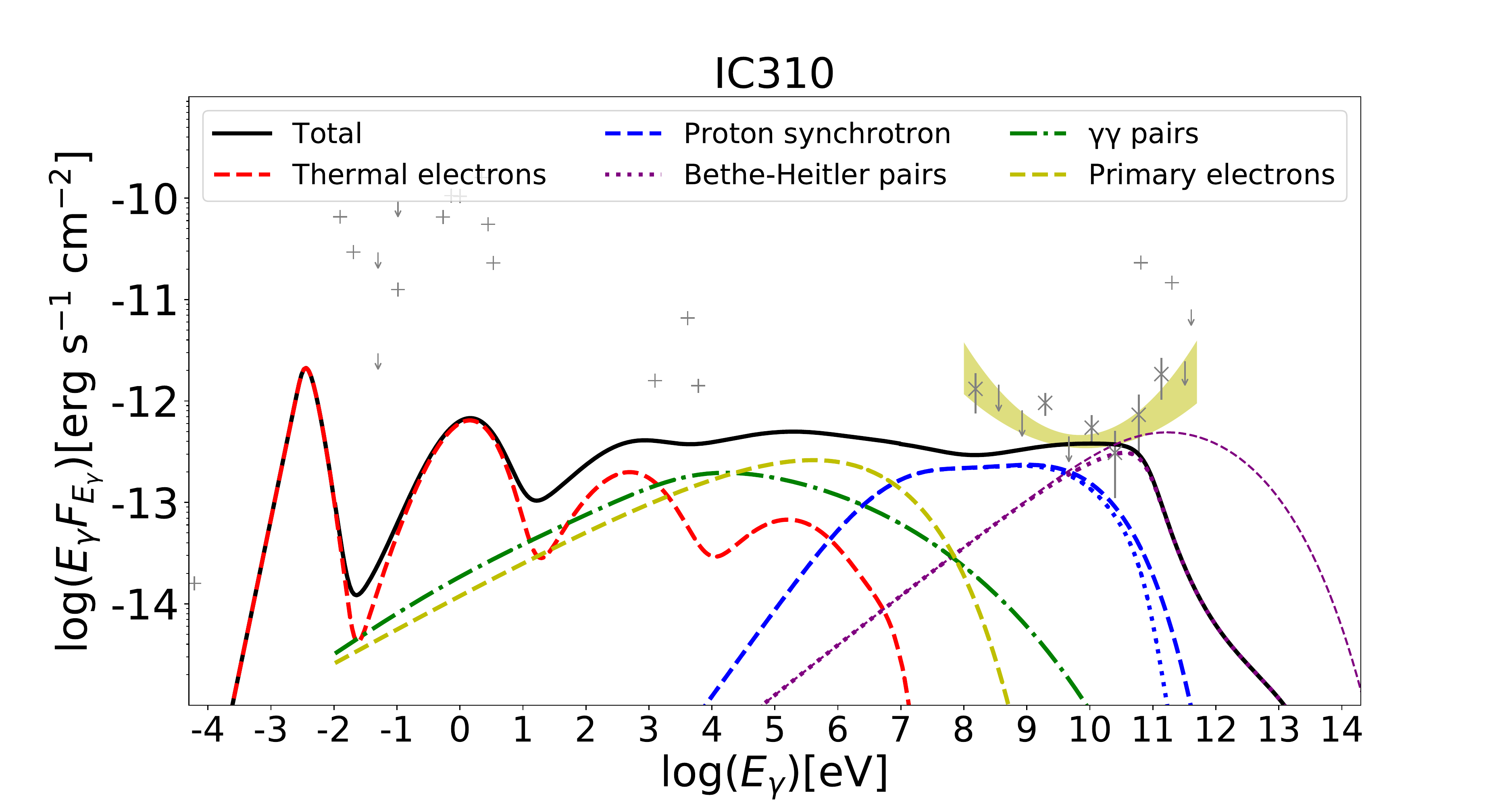}{0.33\textwidth}{}
            \fig{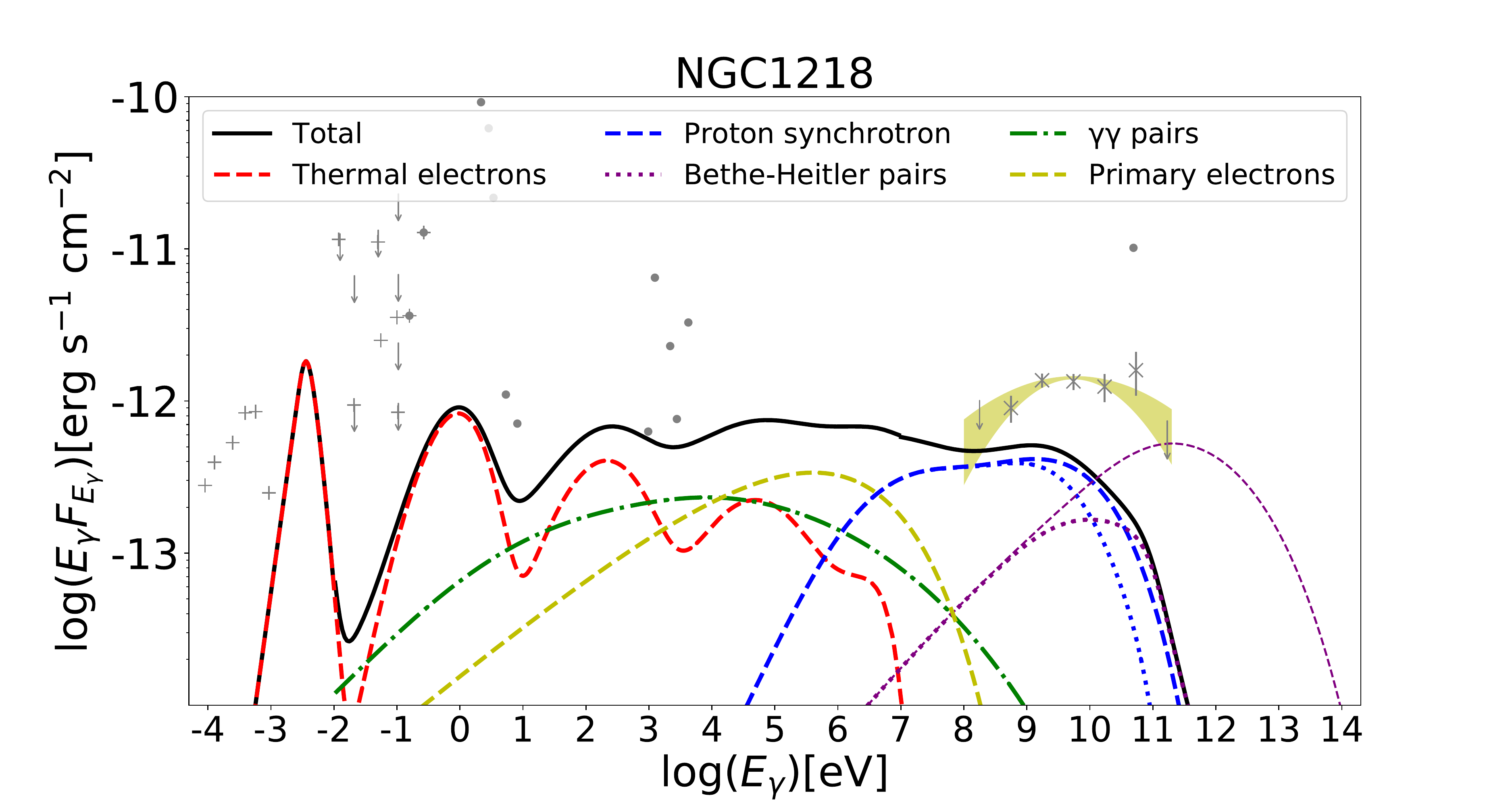}{0.33\textwidth}{}
            \fig{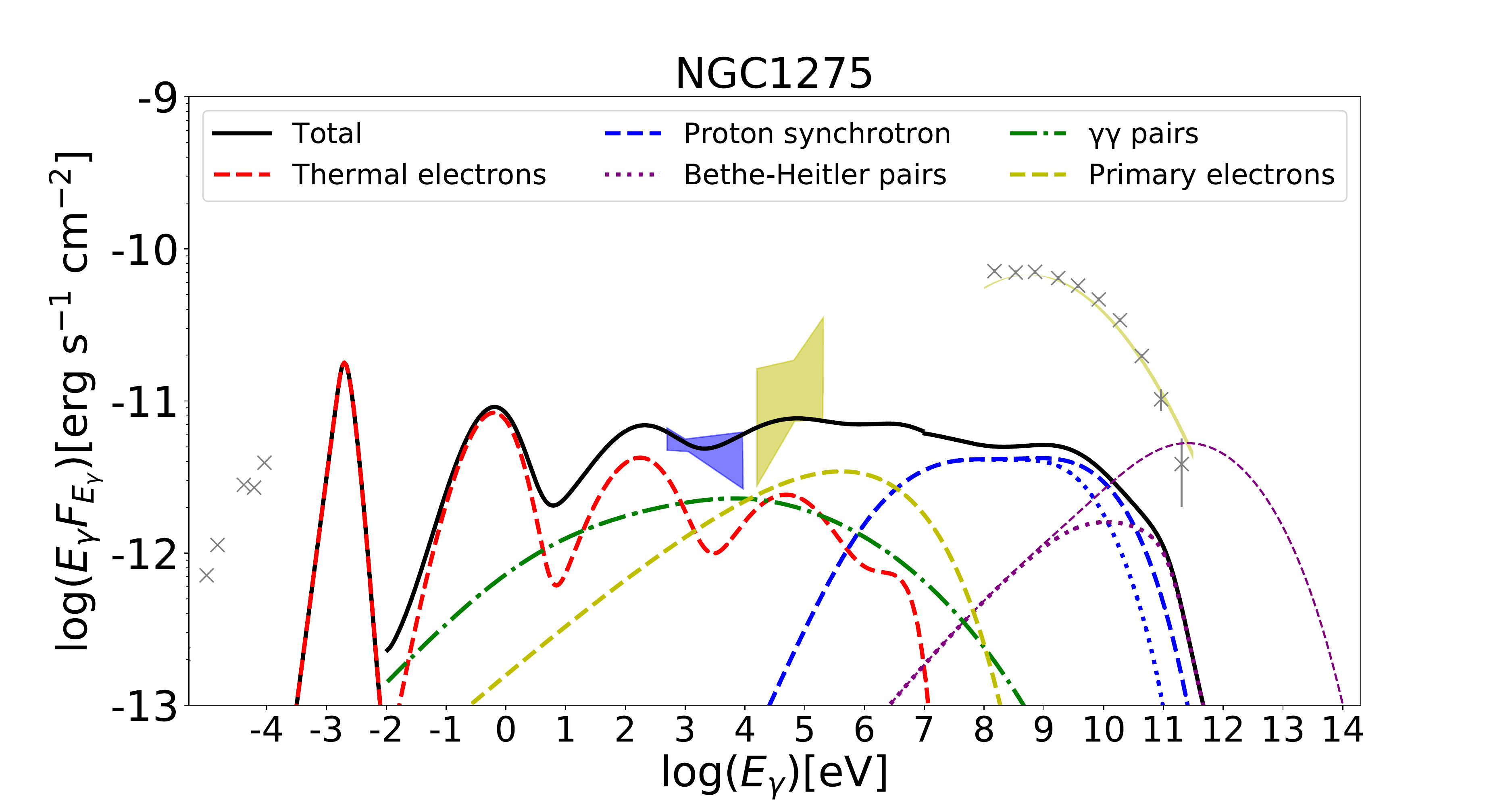}{0.33\textwidth}{}
          }
\gridline{\fig{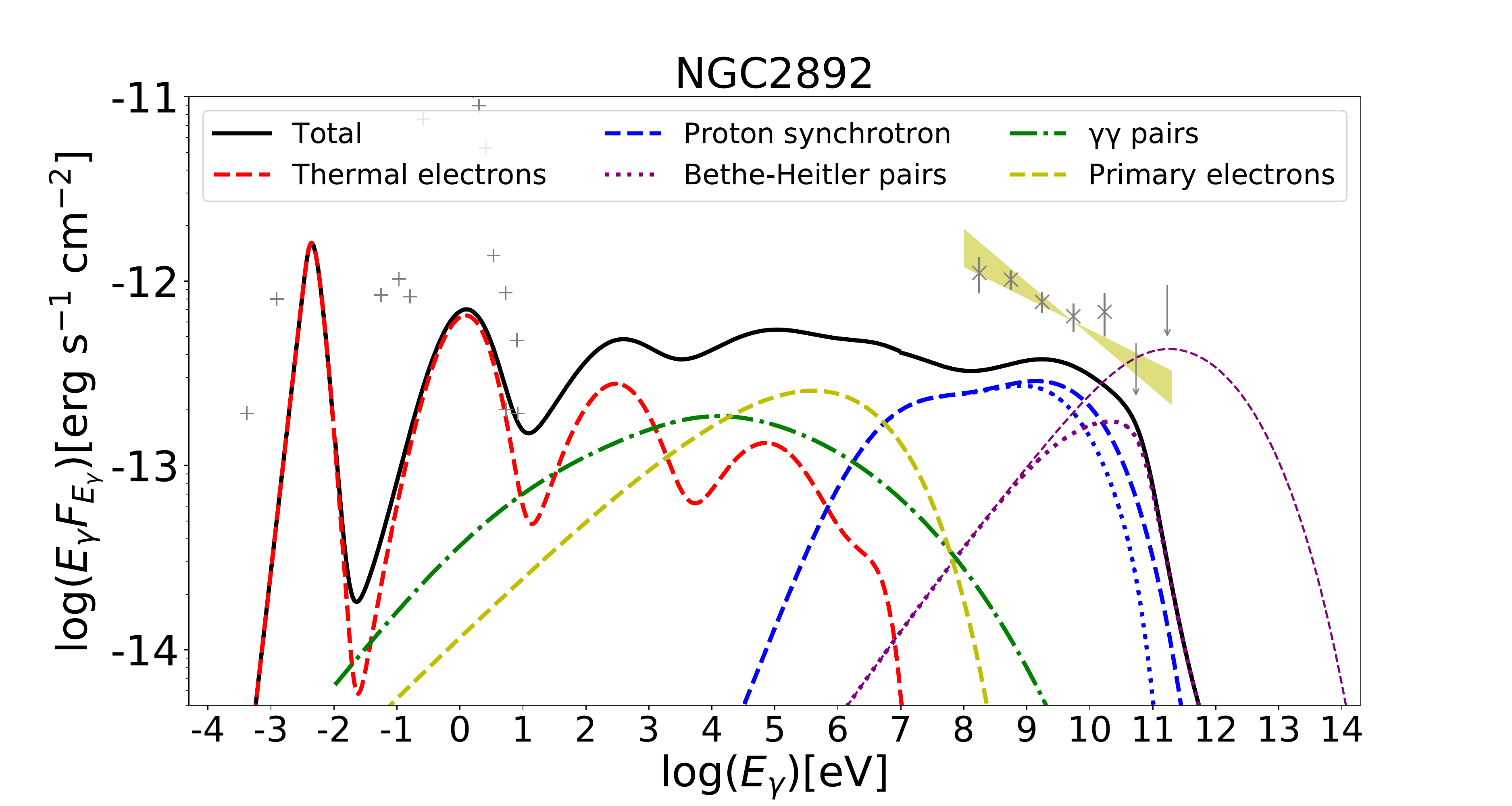}{0.33\textwidth}{}
            \fig{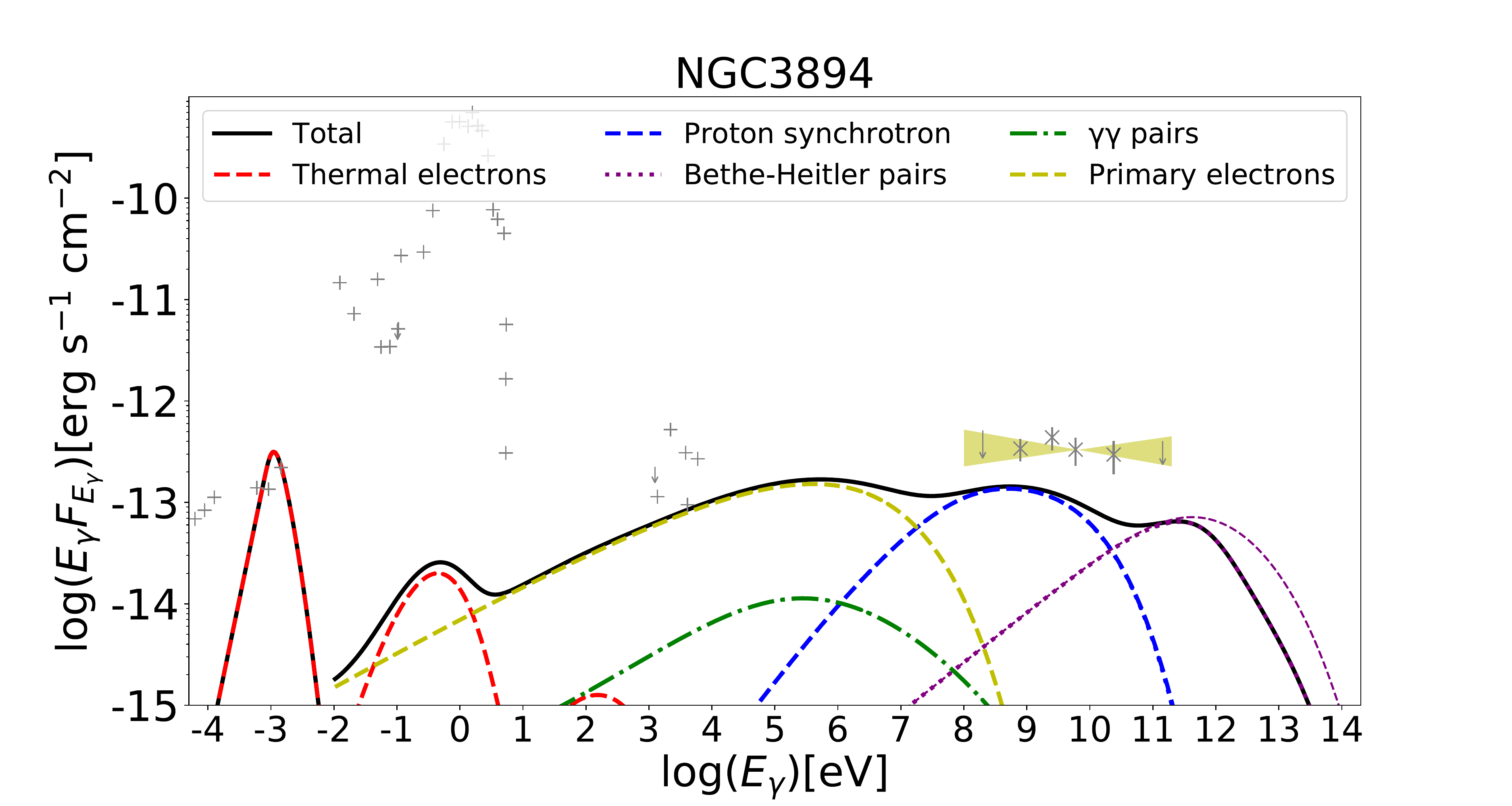}{0.33\textwidth}{}
            \fig{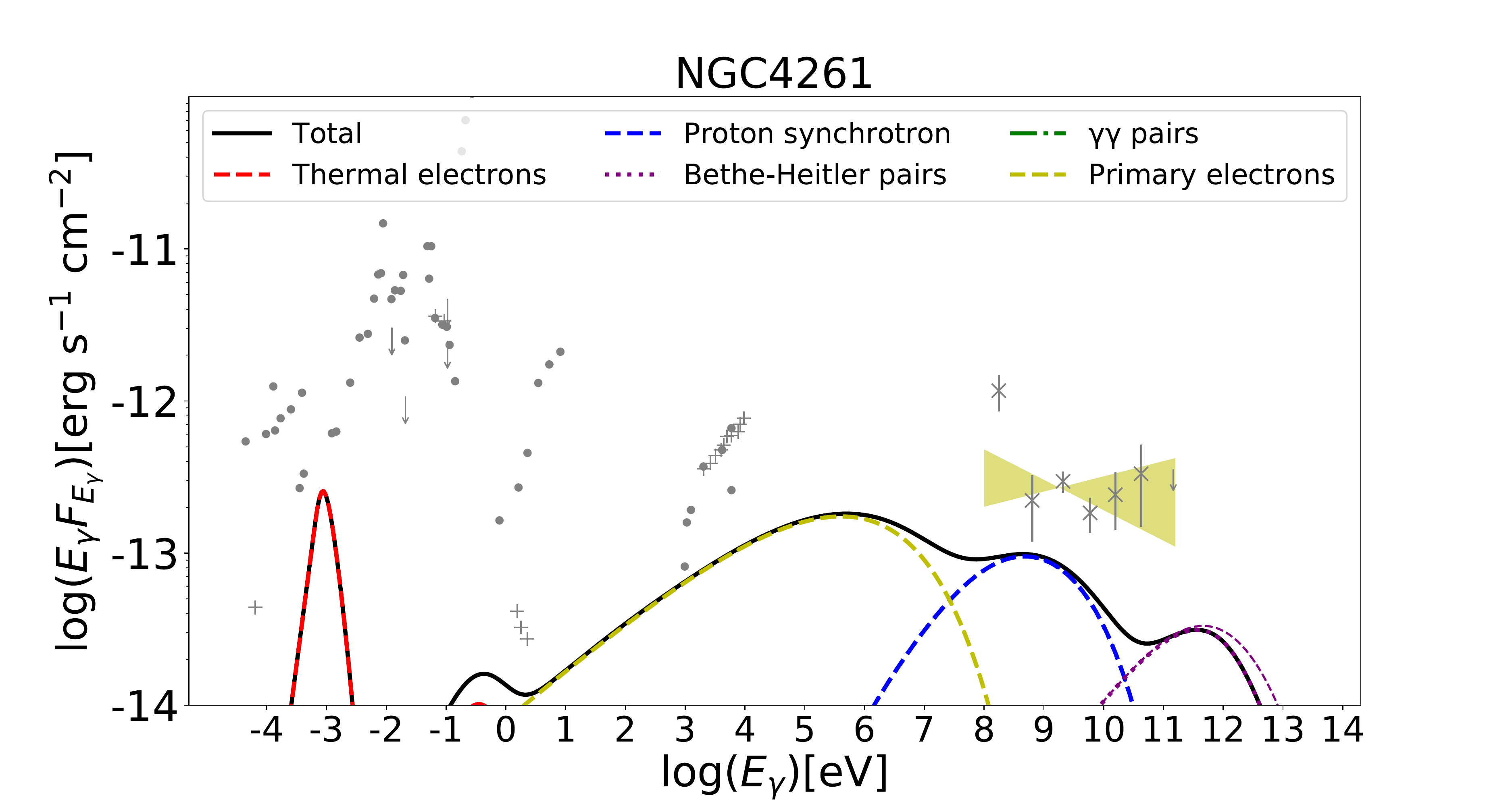}{0.33\textwidth}{}
            }
\gridline{\fig{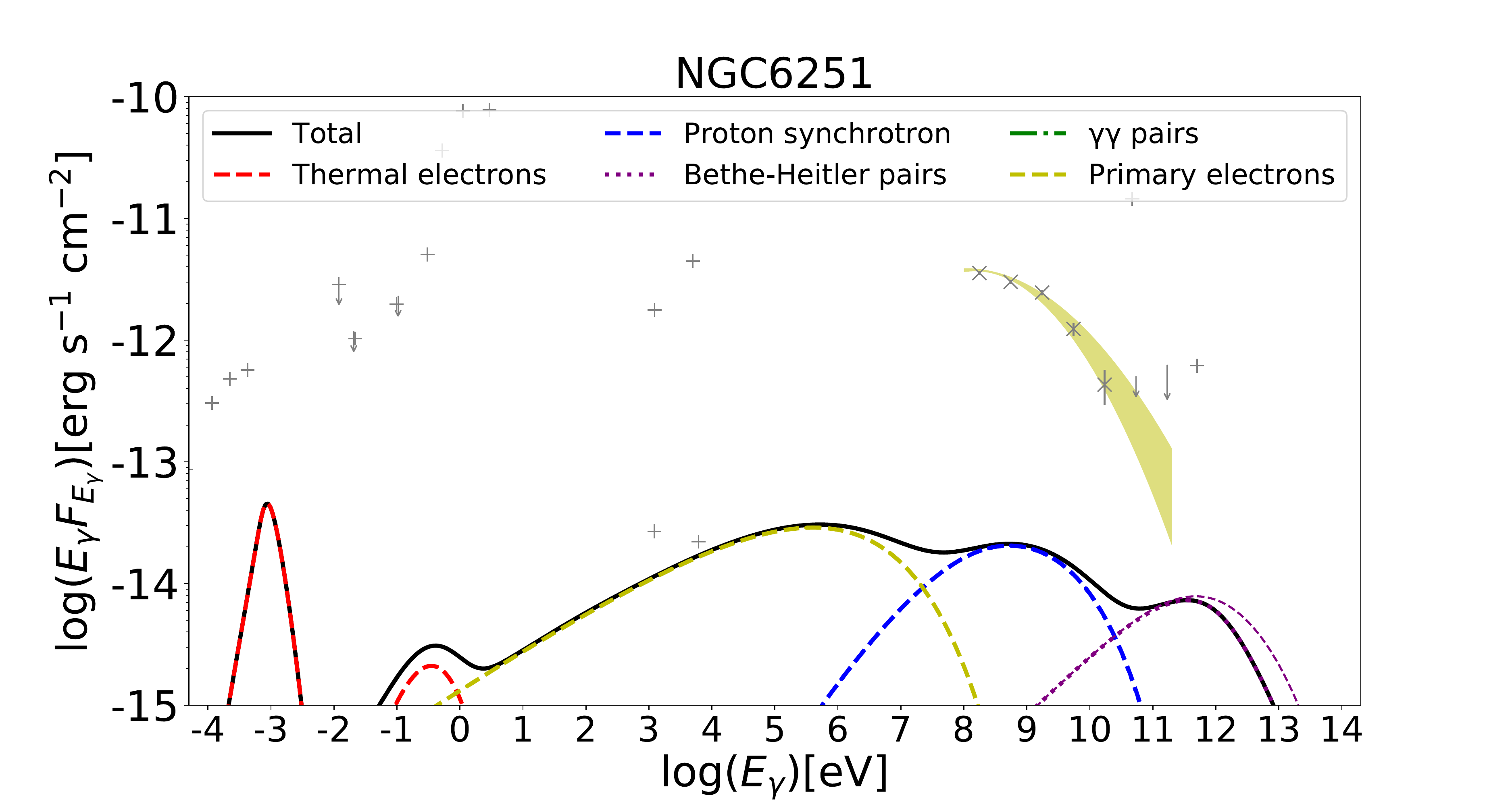}{0.33\textwidth}{}
        \fig{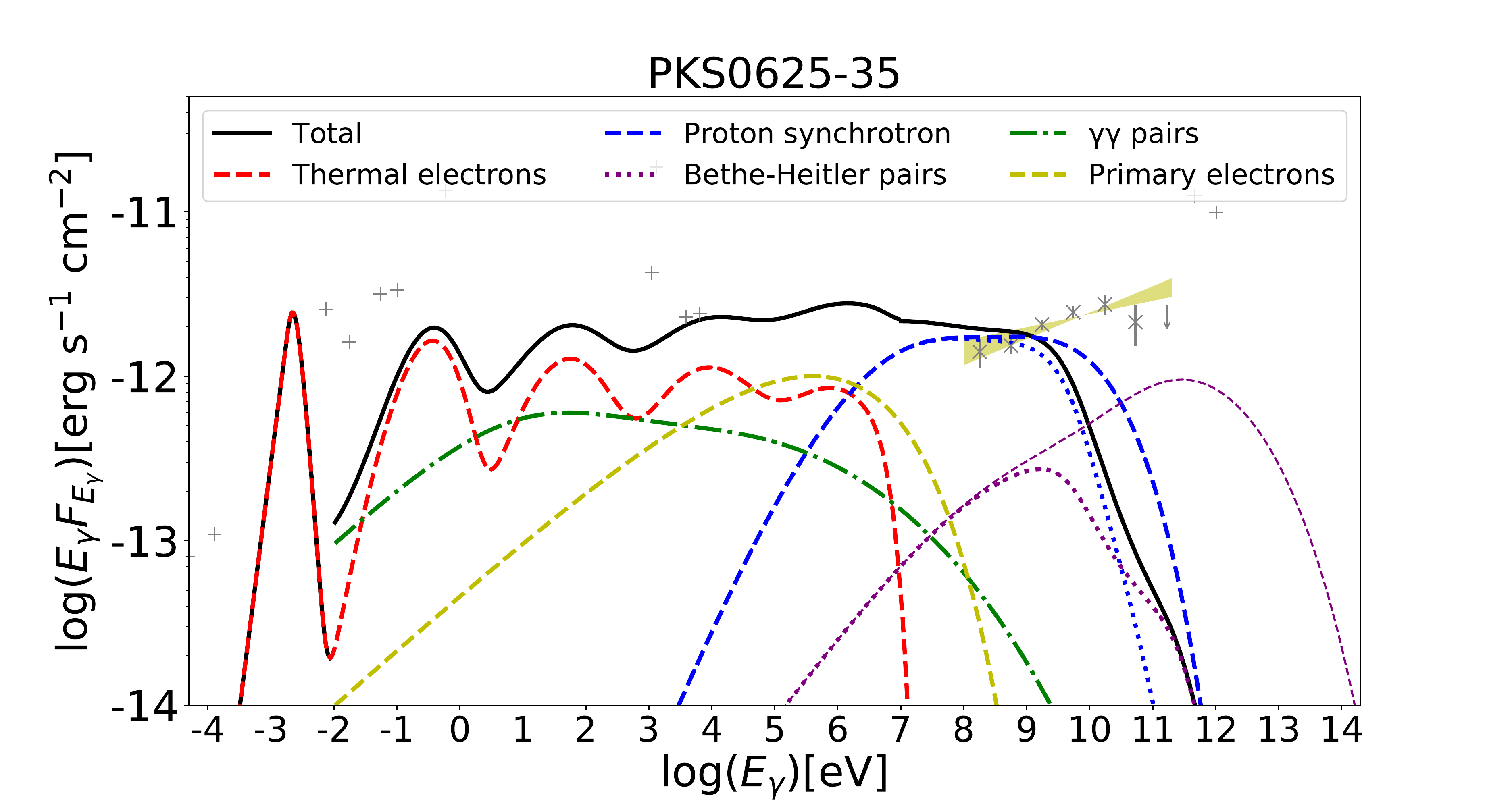}{0.33\textwidth}{}
        \fig{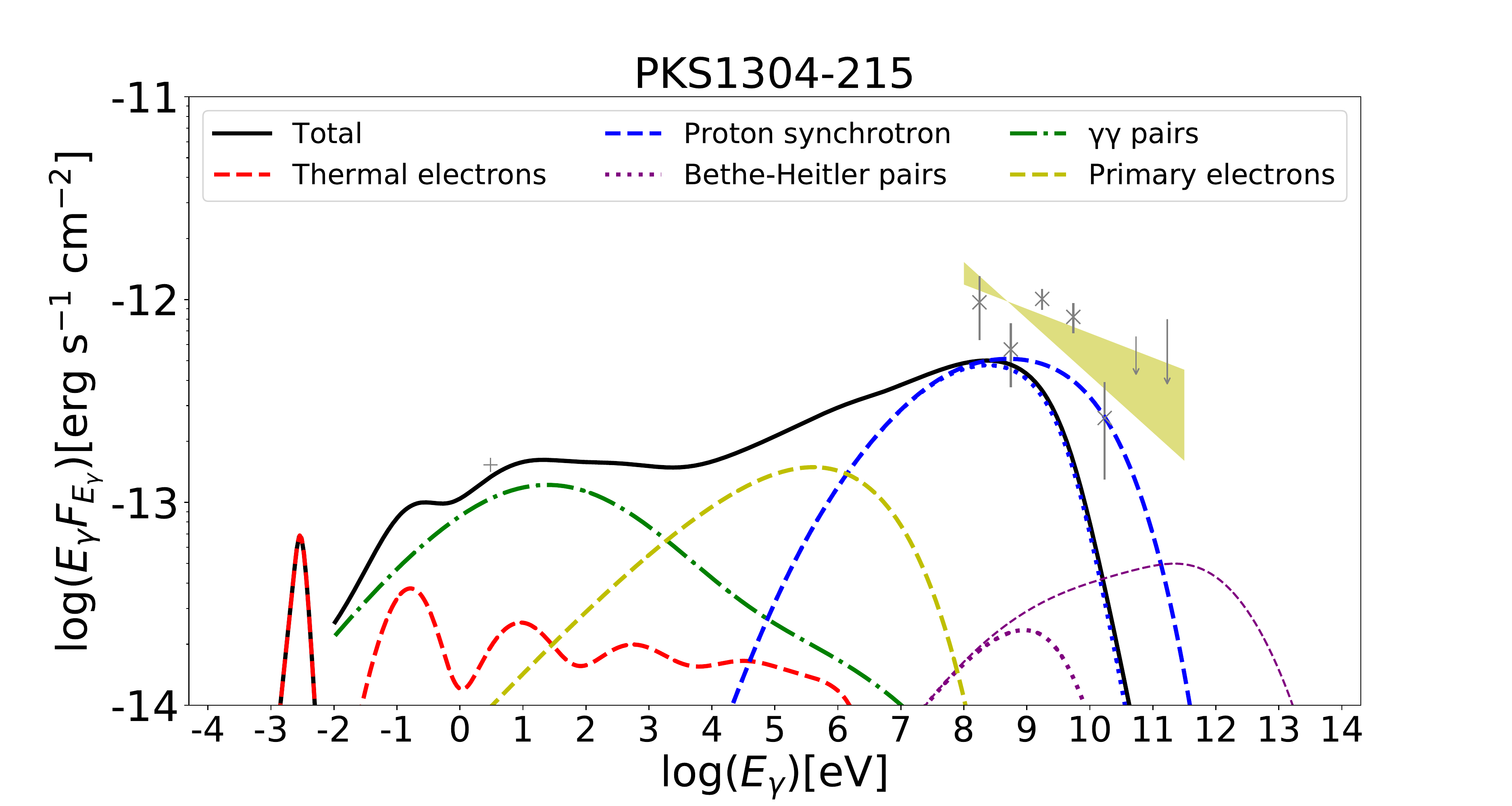}{0.33\textwidth}{}
        }
\gridline{
        \fig{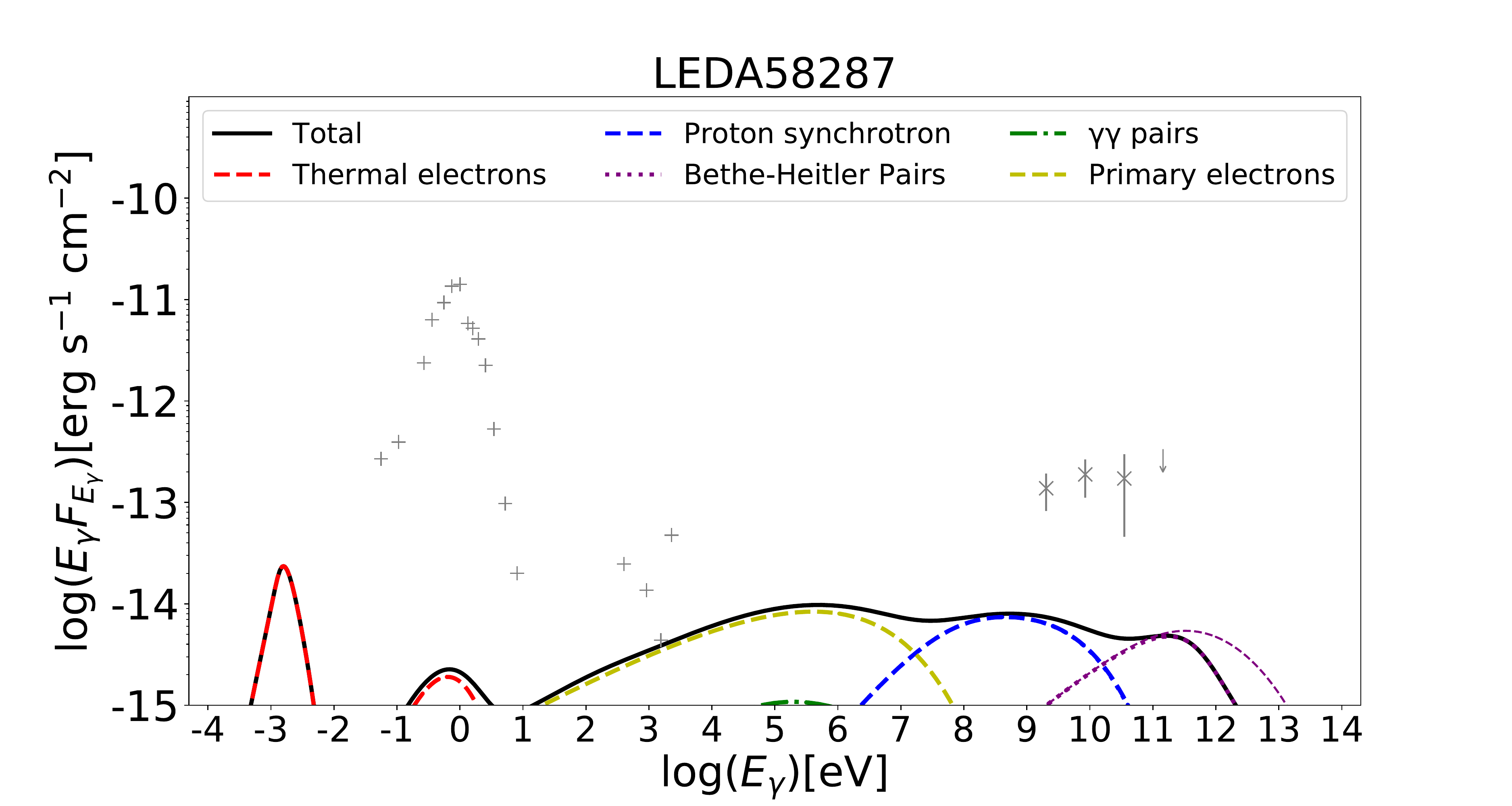}{0.33\textwidth}{}
        \fig{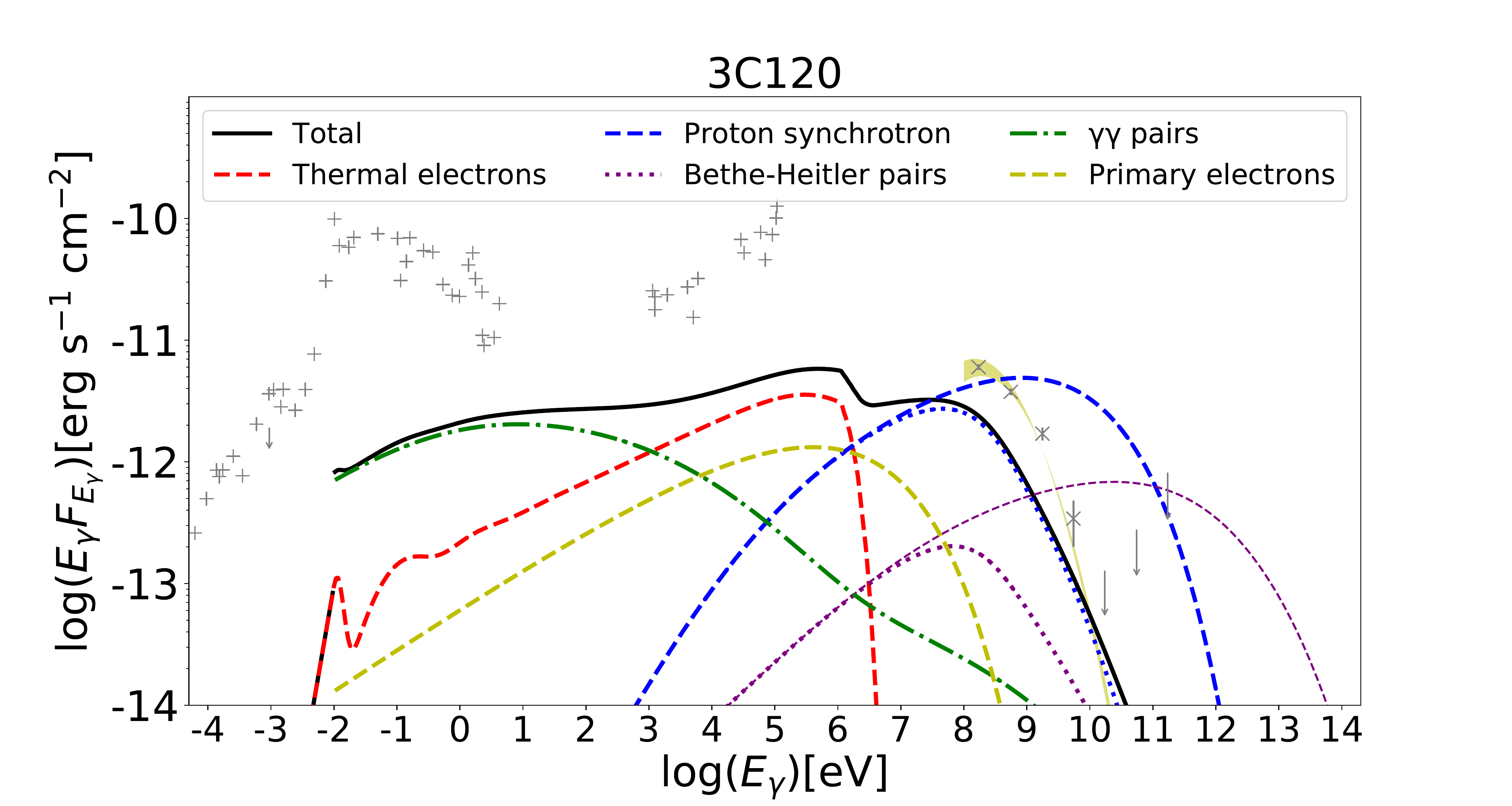}{0.33\textwidth}{}
         \fig{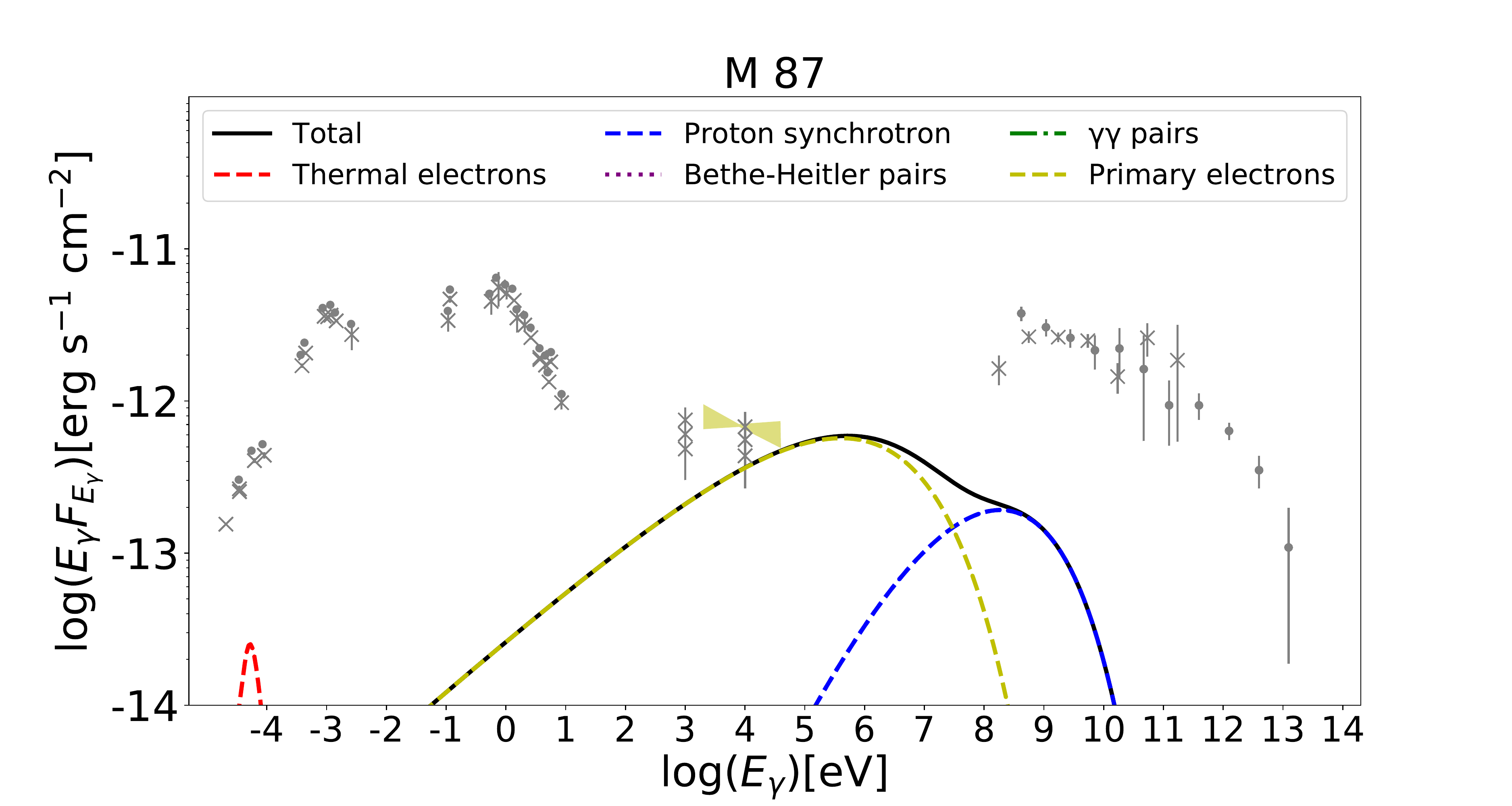}{0.33\textwidth}{}
            }
 \caption{Same as Figure \ref{fig_bd}, but with the electron heating rate given by \cite{Chael2018}. Data points are taken from \cite{MAGIC2020},\cite{Prieto2016},\cite{Wong2017}, and \cite{Benkhali2019} for M87.}
 \label{fig_bd_chael}
\end{figure}

We show the photon spectra of the various radio galaxies with the prescription given by \cite{Chael2018}. The photon spectra of the Excellent, Good, and Bad objects are shown in Figure \ref{fig_vg_chael}, Figure \ref{fig_gd_chael}, and Figure \ref{fig_bd_chael}, respectively. We tabulated the resulting quantities in Table \ref{tb_chael}.

\begin{deluxetable}{ccccccc}
\tablenum{6}
\tablecaption{Quantities and classification with the electron heating rate given by \cite{Chael2018}.
\label{tb_chael}}
\tablewidth{0pt}
\tablehead{
\colhead{Name} & \colhead{$\dot{m}$} & \colhead{$\chi^2/\nu$} & \colhead{$Q$} & Classification }
\startdata
    NGC 2329 & $7.3 \times 10^{-4}$ & 2.1  & 0.1 & Excellent\\
    Cen A & $3.9 \times 10^{-4}$ & 2.3  & 0.06 & Excellent \\
    LEDA 55267 & $3.8 \times 10^{-2}$ & 2.2 & 0.1  & Good\\
    LEDA 57137 & $1.9 \times 10^{-3}$ & 0.23  & 0.8 & Good\\ 
    NGC 315 & $1.7 \times 10^{-4}$ & 1.6 & 0.14 & Good \\
    IC 310 & $8.5 \times 10^{-4}$ & 3.2 & 0.0075 & Bad \\
    NGC 1218 & $2.0 \times 10^{-3}$ & 23  & $2.9 \times 10^{-19}$ & Bad \\
    NGC 1275 & $1.7 \times 10^{-3}$ & 843  & 0.0 & Bad \\
    NGC 2892 & $1.7 \times 10^{-3}$ & 16  & $1.6 \times 10^{-13}$ & Bad \\
    NGC 3894 & $1.7 \times 10^{-4}$ & 7.6  & $4.2 \times 10^{-5}$ & Bad \\
    NGC 4261 & $1.1 \times 10^{-4}$ & 9.1  & $2.5\times10^{-7}$ & Bad \\
    NGC 6251 & $1.3 \times 10^{-4}$ & 481 & 0.0 & Bad \\
    LEDA 58287 & $2.2 \times 10^{-4}$ & 6.4 & 0.0016 & Bad \\
    PKS 0625-35 & $4.9 \times 10^{-3}$ & 25 & $1.7 \times 10^{-25}$ & Bad \\
    PKS 1304-215 & $1.1 \times 10^{-2}$ & 14 & $9.4 \times 10^{-12}$ & Bad \\
    3C 120 & $1.0 \times 10^{-1}$ & 97 & $1.1 \times 10^{-62}$ & Bad \\
    M87 & $1.6 \times 10^{-5}$ & 44 & $6.1 \times 10^{-148}$ & Bad \\
\enddata
\end{deluxetable}


\bibliography{sample631}{}
\bibliographystyle{aasjournal}



\end{document}